\documentclass[]{aa}
\usepackage[varg]{txfonts}
\usepackage{graphics}
\usepackage{float}
\def\cc{\,{\rm cm^{-3}}} 
\def\cm2{\,{\rm cm^{-2}}}

\def\kms{\,{\rm {km\,s^{-1}}}} 
\def\kkms{\,{\rm {K\,km\,s^{-1}}}}

\def\co{\,{\rm ^{12}CO}} 
\def\thirco{\,{\rm ^{13}CO}} 
\def\h2{\,{\rm H_2}} 
 
\def\C34S{\,{\rm C^{34}S}} 

\def\ci{\,{\rm [C I]}}

\def\cii{\,{\rm [C II]}}

\def\etal{et\,al.\ }
\def\aua{{\it A\&A} } 
 
\def\apj{{\it ApJ} } 
 
\def\apjs{{\it ApJS} } 
\def\apjl{{\it ApJL} } 
\def\araa{{\it ARAA} } 
\def\mnras{{\it MNRAS} }

\def\pasp{{\it PASP} } 
\begin{document} 

\title{Central molecular zones in galaxies: \\
$\thirco$(6-5) and molecular gas conditions in bright nearby galaxies}

\author{F.P. Israel\inst{1} 
  \and   R. G\"usten\inst{2}
  \and   A. Lundgren\inst{3}
          } 

\offprints{F.P. Israel} 
 
\institute{Sterrewacht Leiden, P.O. Box 9513, 2300 RA Leiden, the Netherlands
  \and       Max-Planck-Institut f\"ur Radioastronomie, 
             Auf dem H\"ugel 69, 53121 Bonn, Germany
  \and       Aix Marseille Universit\'e, CNRS, LAM, Marseille, F-13388, France 
           }
 
\date{Received ????; accepted ????}

\abstract{

  This paper summarizes all presently available $J_{upp}\geq5$
  $\thirco$ and accompanying $\co$ measurements of galaxy centers
  including new $J$=6-5 $\thirco$ and $\co$ observations of eleven
  galaxies with the Atacama Pathfinder EXperiment (APEX) telescope and
  also $Herschel$ high-$J$ measurements of both species in five galaxies.
  The observed $J$=6-5/$J$=1-0 $\co$ integrated temperature ratios
  range from 0.10 to 0.45 in matching beams.  Multi-aperture data
  indicate that the emission of $\thirco$(6-5) is more centrally
  concentrated than that of $\co$(6-5).  The intensities of $\co$(6-5)
  suggest a correlation with those of HCO$^{+}$ but not with those of
  HCN.  The new data are essential in refining and constraining the
  parameters of the observed galaxy center molecular gas in a simple
  two-phase model to approximate its complex multi-phase structure.
  In all galaxies except the Seyfert galaxy NGC~1068, high-$J$ emission
  from the center is dominated by a dense ($n\sim10^{5}\cc$) and
  relatively cool (20-60 K) high-pressure gas. In contrast, the
  low-$J$ lines are dominated by low-pressure gas of a moderate density
  ($n\sim10^3\cc$) and more elevated temperature (60-150 K) in most
  galaxies. The three exceptions with significant high-pressure gas
  contributions to the low-$J$ emission are all associated with active
  central star formation. } 

\maketitle

\section{Introduction}

\begin{table}
\begin{center}
{\small %
\caption[]{\label{crit}Line excitation.}
\begin{tabular}{lccccc}
\noalign{\smallskip}     
\hline
\noalign{\smallskip}
            &\multicolumn{2}{c}{Frequency} &E$_{u}$/k$^{a}$ &$n^{b}_{crit}$ \\
$J$         &\multicolumn{2}{c}{(GHz)}     & (K)           & ($\cc$)    \\
  (1)       & (2)        &   (3)           & (4)           & (5)        \\
\noalign{\smallskip}      
\hline
\noalign{\smallskip} 
            & $\thirco$         & $\co$    &       &      \\
\noalign{\smallskip} 
\hline
\noalign{\smallskip}  
1-0         & 110.201           & 115.271  &   5.5 & 2e3  \\
2-1         & 220.399           & 230.538  &  16.6 & 1e4  \\
3-2         & 330.588           & 345.796  &  33.2 & 4e4  \\
4-3         & 440.765           & 461.041  &  55.3 & 9e4  \\
5-4         & 550.926           & 576.278  &  83.0 & 2e5  \\
6-5         & 661.067           & 691.473  & 116.2 & 3e5  \\ 
7-6         & 771.184           & 806.652  & 154.9 & 5e5  \\
8-7         & 881.273           & 921.800  & 199.1 & 6e5  \\
\noalign{\smallskip}      
\hline
\noalign{\smallskip}
            & HCN               & HCO$^+$  &       &      \\             
\noalign{\smallskip} 
\hline
\noalign{\smallskip}  
1-0         &  88.632           &  ...     &  4.3  & 2e5  \\
1-0         &  ...              &  89.189  &  4.3  & 3e4  \\
3-2         & 265.886           &  ...     & 25.5  & 4e6  \\
3-2         &  ...              & 267.558  & 25.5  & 8e5  \\
\noalign{\smallskip} 
\hline
\end{tabular} 
}%
\end{center} 
Notes: a. Jansen (1995); Sch\"oier $\etal$ (2005); b. Calculated for
kinetic temperatures $T_{\rm kin}(\rm CO)$ = 100 K in the optically thin limit
ignoring radiative trapping; Carilli \& Walter (2013).
\end{table}   

Late-type galaxies frequently contain conspicuous central
concentrations of molecular gas. These may play an important role in
galaxy evolution when they serve as the reservoirs that feed
super-massive black holes, circumnuclear star formation, and massive
gas outflows. Various $\co$ ladder surveys, in particular those conducted
with the $Herschel$ Space Observatory, unambiguously point to the
simultaneous presence of both low-pressure and high-pressure gas in
these reservoirs (Mashian $\etal$ 2015, Kamenetzky et al. 2016, 2017,
Lu $\etal$ 2017, Crocker et al. 2019), requiring a multi-phase
analysis.

Table\,\ref{crit} illustrates how different molecular line
transitions, in principle, can be used to determine molecular gas
temperatures and densities. Yhe values in that table were, however,
calculated ignoring radiative trapping and assuming optically thin
emission, whereas the $\co$, HCN, and HCO$^+$ transitions are
optically thick. Significant emission occurs well below the critical
density, at densities lower by one or two orders of magnitude
(cf. Shirley 2015).  Nevertheless, the table provides useful upper
limits to the temperature and the density that can be deduced from a
transition-limited survey. For instance, $\co$ $J$=2-1/$J$=1-0
intensity ratios distinguish kinetic temperatures omcreasingly poorly
above ~20 K, and $\co$ ladders up to $J$=4-3 fail to make meaningful
distinctions between temperatures above ~100 K.

A complication is the degeneracy of optically thick $\co$ ladders with
respect to the kinetic temperature, volume density and column density
(hence mass) of the gas. A striking illustration of their failure to
differentiate between even the very different environmental conditions
in NGC~6240 and Mrk~231 is provided by Meijerink $\etal$
(2013).  Likewise, Weiss $\etal$ (2007) found equally good fits to the
$\co$ ladders of luminous galaxies but they could not resolve the
temperature-density ambiguity. Additional information preferably in
the form of optically thin emission from species such as $\thirco$ is
required to alleviate the degeneracy (Bayet $\etal$ 2006, Israel 2020,
hereafter Paper I).  Depending on the available data, two or three gas
phases can be modeled. For most purposes, a two-phase analysis
suffices as it can be made to fit most of the available observations
(e.g., the cases of M~82 and NGC~253 discussed by Loenen et al. 2010
and by Rosenberg $\etal$ 2014).

The presently available data on galaxy centers do not constrain the
relative amounts of low-pressure and high-pressure gas equally
well. In Paper I, we presented a systematical probe of the physical
condition of the molecular gas with ground-based surveys of both $\co$
and $\thirco$ in transitions up to $J_{upper}=4$ and found densities
between $10^2$ cm$^{-3}$ and $\geq$10$^4$ cm$^{-3}$ and temperatures
ranging from $\sim$30 K to $\geq$100 K.  The elevated gas
temperatures, increased turbulence, and higher metallicities that
characterize galaxy centers cause a systematic overestimate of the
molecular hydrogen amounts by traditional methods. Instead, the
so-called $X$-factor relating CO intensities to H$_{2}$
column densities is an order of magnitude lower than the ``standard''
value in galaxy centers, including the Milky Way center.
  
These low-$J$ transitions are particularly sensitive to molecular gas
of a modest density (cf. Table\,\ref{crit}) and constrain the column
density and mass of the low-pressure gas relatively well, even though
the available line intensities usually do not fully constrain even a
two-phase model. In particular, they do not adequately sample the
temperatures and densities at the high-pressure end which are much
more sensitive to feedback from active-galaxy nuclei (AGN) and from
starburst activity. This requires additional surveys of the higher-$J$
$\co$ transitions such as those provided by the $Herschel$ Spectral and
Photometric Imaging Receiver (SPIRE) and Photoconductor Array Camera
and Spectrometer (PACS) that cover $\co$ transitions $J_{upp}\geq 4$
in a large number of galaxies (e.g., Mashian $\etal$ 2015, Rosenberg
et al. 2015, Kamenetzky et al. 2016). Such observations were attempted
with the $Herschel$ Heterodyne Instrument for the Far-Infrared (HIFI)
overlapping the few cases where SPIRE sensitivities did also allow the
determination of $\thirco$ line fluxes.

High-frequency observations of extragalactic $\thirco$ lines are also
feasible with ground-based equipment but only at the high-elevation
facilities in Hawaii and the Chilean Andes. High atmospheric opacities
prevent ground based observation of the $J$=4-3 and $J$=5-4 $\thirco$
transitions. The $\thirco$ line intensities, already low in almost all
galaxies, further decrease with increasing $J$ level. This leaves the
$J$=6-5 $\thirco$ line as the most practical choice to sample the
high-excitation gas in galaxy centers from the ground.

\section{Galaxy sample}

The sample considered here includes the few galaxies with $Herschel$
detections of $\thirco$ in $J_{upp}\geq5$ transitions. These concern
fluxes extracted from SPIRE spectra covering a great spectral range
with low spectral resolution and from targeted Heterodyne Instrument
for the Far-Infrared (HIFI) observations with much higher spectral
resolution resolving the line profiles. Although SPIRE detected many
galaxies in various $\co$ transitions, the weak $\thirco$ lines were
unambiguously detected only in the brightest galaxies on the celestial
sky mostly in guaranteed observing time.

The $\co(6-5)$ line was readily observed from the ground, first in the
bright galaxies accessible from Hawaii, notably M~82, NGC~253, and
IC~342 (Harris $\etal$ 1991; Ward $\etal$ 2003, Seaquist $\etal$ 2006)
and in some red-shifted luminous galaxies from lower-elevation sites
(cf. Weiss \etal 2007). Less luminous closer galaxies followed
(e.g., Bayet$\etal$ 2004, 2006), but the ground-based detection of the
$\thirco(6-5)$ line in NGC~253 (Hailey-Dunsheath $\etal$ 2008) so far
stood alone. The development of sensitive high-frequency receivers for
use in the southern hemisphere provided the opportunity to change this
situation.  Inspection of the $J$=6-5 $\co$ data in the $Herschel$
archive yielded a limited number of galaxies bright enough to attempt
$J$=6-5 $\thirco$ detection from the southern hemisphere without the
need for prohibitively long integration times. The sample selected for
new observations is listed in Table\,\ref{sample}. It includes six
galaxies with starburst centers, three have AGN centers, one is the
Luminous InfraRed Galaxy merger (LIRG) NGC 6240, and one has a mixed
AGN-starburst center (NGC 1365). The two northern galaxies bright
enough to have literature mid-$J$ $\thirco$ intensities (M~82 and
IC~342) have been included for completeness sake.

\begin{table}
\begin{center}
{\small %
\caption[]{\label{sample}Sample galaxies.}
\begin{tabular}{lrrrr}
  \noalign{\smallskip}     
    \hline
  \noalign{\smallskip} 
Name   & R.A. (2000) & Dec. 2000)    & $V_{\rm LSR}$  & $D$ \\
       &  h m s      &$\circ$ $'$ $"$& $\kms$       & Mpc  \\
(1)    & (2)         &     (3)       & (4)          & (5) \\
  \noalign{\smallskip}      
    \hline
  \noalign{\smallskip} 
\multicolumn{5}{c}{Newly observed galaxies} \\
  \noalign{\smallskip}
    \hline
  \noalign{\smallskip}
N253   & 00:47:33.1 & -25:17:18 &   245 &   3.4 \\
N613   & 01:34:18.2 & -29:25:06 &  1480 &  19.7 \\
N660   & 01:43:02.4 & +13:38:42 &   843 &  12.2 \\
N1068* & 02:42:40.7 & -00:00:48 &  1135 &  15.2 \\
N1097* & 02:46:19.0 & -30:16:30 &  1270 &  16.5 \\
N1365* & 03:33:36.4 & -36:08:25 &  1635 &  21.5 \\
N1808  & 05:07:42.3 & -37:30:47 &   995 &  12.3 \\
N2559  & 08:17:06.1 & -27:27:21 &  1560 &  21.4 \\
N4945* & 13:05:27.5 & -49:28:06 &   565 &   4.4 \\
N5236  & 13:37:00.9 & -29:51:56 &   515 &   4.0 \\
N6240  & 16:52:58.9 & +02:24:03 &  7304 & 116.4 \\ 
  \noalign{\smallskip}      
    \hline
  \noalign{\smallskip}  
\multicolumn{5}{c}{Galaxies from the literature} \\
  \noalign{\smallskip}
    \hline
  \noalign{\smallskip}
IC342 & 03:46:48.5 & +68:05:47 &    35 &   3.5 \\    
N3034 & 09:55:52.7 & +69:40:46 &   275 &   4.0 \\
\noalign{\smallskip}     
\hline
\end{tabular}
}%
\end{center} 
Note: Active galaxy nuclei (AGNs) are marked by an asterisk.
\end{table}   

\begin{table}
\begin{center}
{\small %
  \caption[]{\label{results}Results of APEX observations.}
\begin{tabular}{lcccc}
\noalign{\smallskip}     
\hline
\noalign{\smallskip} 
Name    & $\int{T_{mb}}$d$v$ & $T_{mb}$ & $\int{T_{mb}}$d$v$ & $T_{mb}$ \\
        & ($\kkms$)         & (mK)    & ($\kkms$)          & (mK)    \\
(1)     & (2)               &    (3)  & (4)                & (5) \\
\noalign{\smallskip}      
\hline
\noalign{\smallskip}
 & \multicolumn{2}{c}{$\co(6-5)^a$} & \multicolumn{2}{c}{$\thirco(6-5)^a$}\\
\noalign{\smallskip}      
\hline
\noalign{\smallskip}      
N253  & 1000.4$\pm$2.2     & 5903$\pm$177 & 75.21$\pm$1.08 & 475$\pm$15 \\
N613  &   39.9$\pm$1.6     &  160$\pm$19  &  2.65$\pm$0.71 &  12$\pm$3  \\
N660  &   48.1$\pm$5.2$^b$ &  252$\pm$37  &  3.79$\pm$1.33 &  21$\pm$5  \\
N1068 &   96.8$\pm$2.9$^b$ &  444$\pm$51  &  4.46$\pm$0.60 &  21$\pm$3  \\ 
N1097 &   49.2$\pm$4.1$^b$ &   90$\pm$22  &  4.48$\pm$1.44 &  13$\pm$4  \\
N1365 &   65.3$\pm$1.2     &  319$\pm$19  &  4.79$\pm$0.65 &  23$\pm$6  \\
      &   64.7$\pm$4.0$^b$ &  419$\pm$39  &   ...          & ...        \\
N1808 &  111.5$\pm$1.7     &  498$\pm$38  &  4.63$\pm$0.58 &  17$\pm$5  \\
      &  116.4$\pm$7.1$^b$ &  404$\pm$41  &   ...          & ...        \\   
N2559 &   29.8$\pm$1.1     &  158$\pm$23  &  2.17$\pm$0.40 &  13$\pm$4  \\ 
N4945 &  724.0$\pm$6.8     & 2508$\pm$165 & 59.88$\pm$2.88 & 206$\pm$41 \\ 
N5236 &   68.4$\pm$0.6     &  650$\pm$42  &  3.23$\pm$0.81 &  33$\pm$9  \\
N6240 &   54.7$\pm$2.7     &  148$\pm$18  &  2.21$\pm$0.56 &   6$\pm$2  \\
\noalign{\smallskip}     
\hline
\noalign{\smallskip} 
  & \multicolumn{2}{c}{$\co(7-6)^c$} & \multicolumn{2}{c}{$\ci(2-1)^c$}\\
\noalign{\smallskip}     
\hline
\noalign{\smallskip} 
N660  & 46.2$\pm$17.5$^b$ & 691$\pm$252 & 34.3$\pm$18.8 & 333$\pm$219$^b$ \\
N1808 & 22.2$\pm$2.2$^b$  & 161$\pm$16  & 36.1$\pm$6.9  & 150$\pm$40$^b$ \\
N1068 &  ...              &  ...        & 69.0$\pm$8.0  & 335$\pm$57$^b$ \\
N1097 &  ...              &  ...        & 98.9$\pm$34.4 & 295$\pm$46$^b$\\ 
\noalign{\smallskip}
\hline
\end{tabular}
}%
\end{center} 
Notes: a. SEPIA660 results unless noted otherwise; b. CHAMP+ results;
c. binned to the $9"$ SEPIA660 resolution
\end{table}   

\begin{figure*}
  \begin{center}
\begin{minipage}[]{18cm} 
    \resizebox{17.67cm}{!}{\rotatebox{0}{\includegraphics*{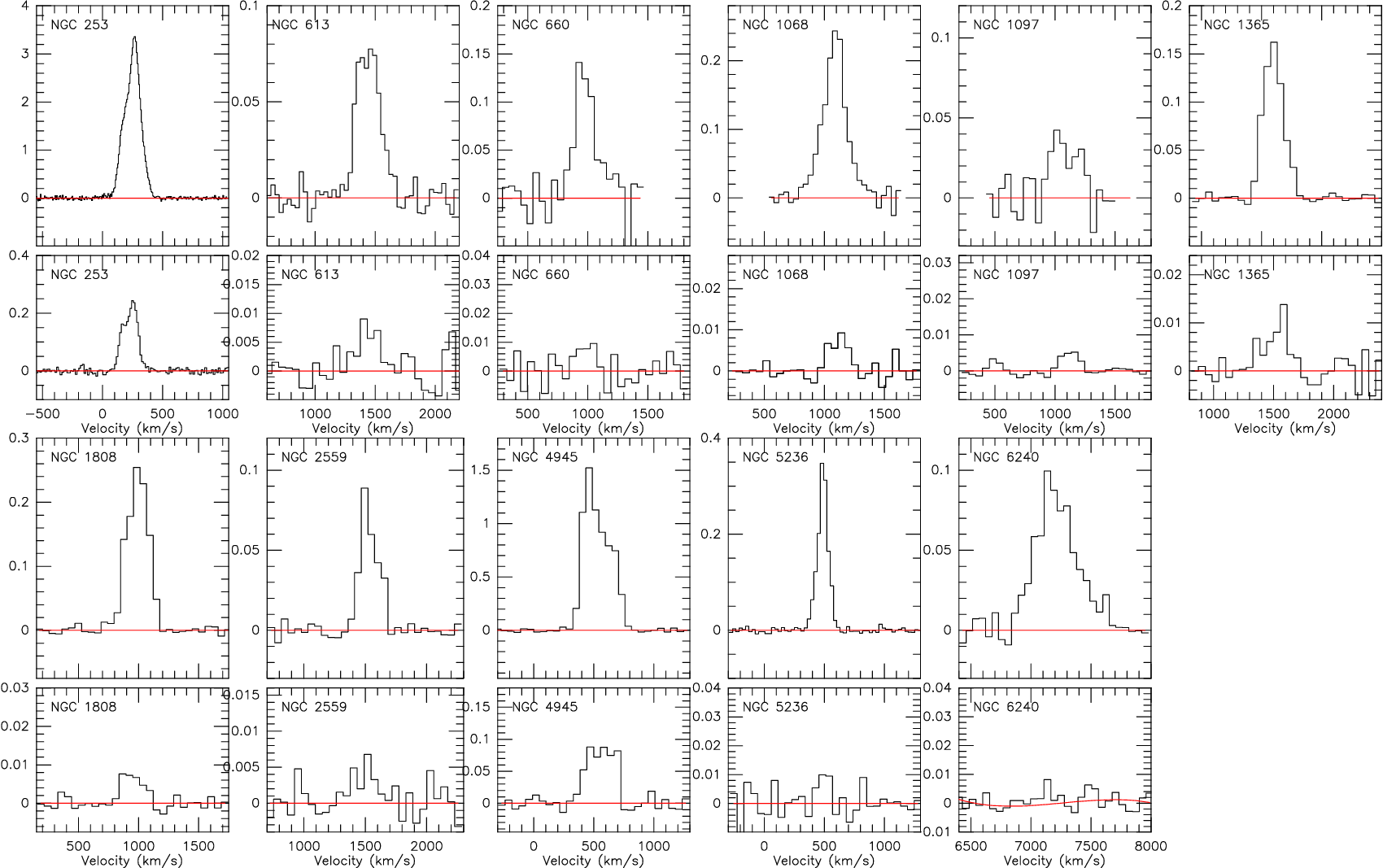}}}
  \end{minipage}
  \caption[] {Observed $J$=6-5 $^{12}$CO (top) and $J$=(6-5) $^{13}$CO
    (bottom ) profiles of galaxy centers. Horizontal scale velocity
    ($V_{LSR}$ in $\kms$, vertical scale observed antenna temperature
     $T_{A^*}$. Note the overall weakness of $^{13}$CO lines.  }
\label{isorat}
\end{center}
\end{figure*}

\begin{figure}
\begin{center}
\begin{minipage}[]{9cm} 
\resizebox{9cm}{!}{\rotatebox{0}{\includegraphics*{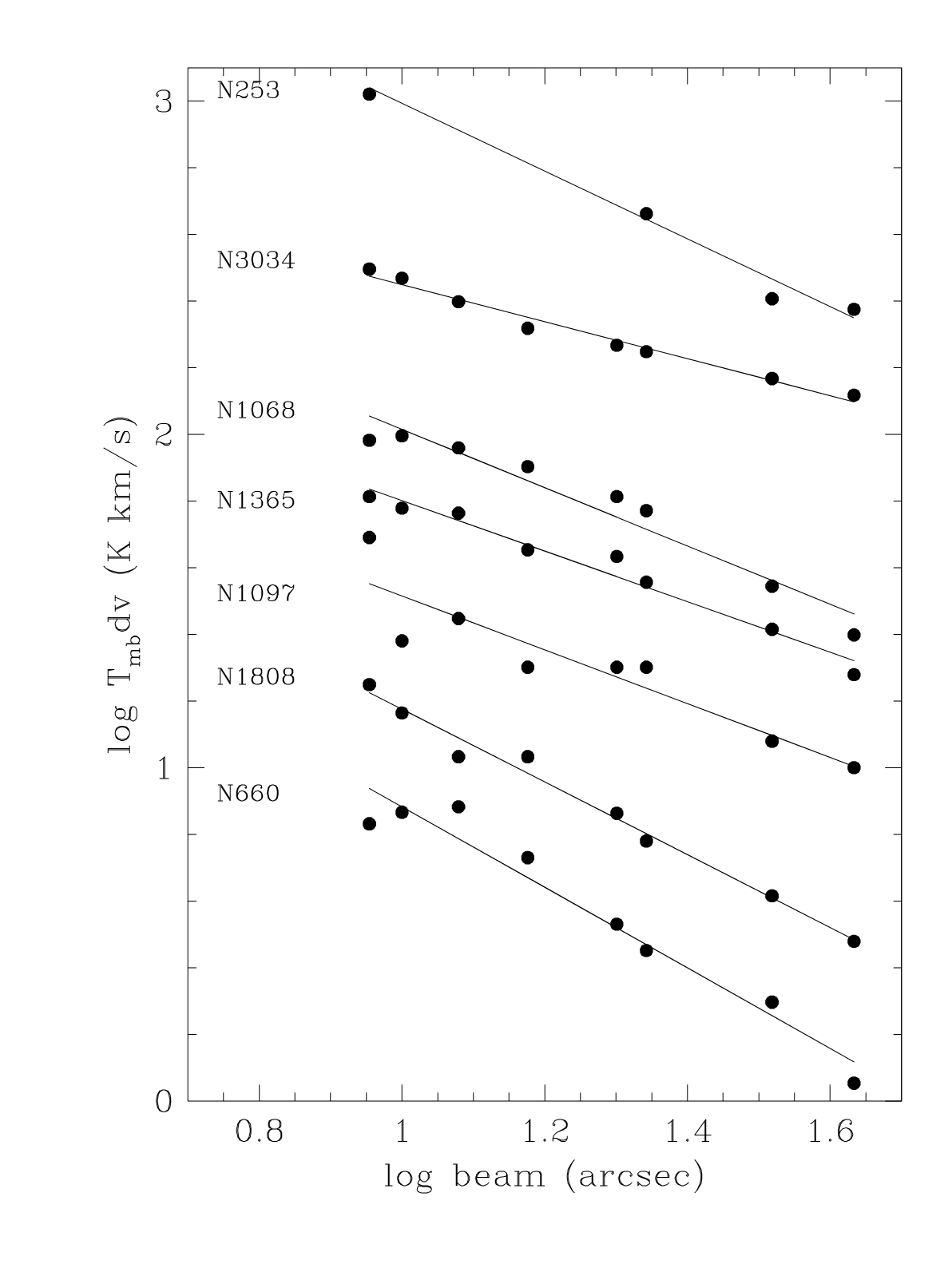}}}
\end{minipage}
\caption[] {Integrated $\co$(6-5) line intensities as a function of
  aperture. For clarity sake, points for NGC~1808 and NGC~660 were
  shifted down by 0.80 and 0.85, respectively. Solid lines are
  least-squares fits to the data. Six galaxies have poor data coverage
  and are not shown. NGC~613, IC~342, and NGC~2559 have a single
  data point each. NGC~4945, NGC~5236, and NGC~6240 have only two data
  points each. }
\label{beamfig}
\end{center}
\end{figure}

\section{Observations}

The observations are part of two separate programs with the Atacama
Pathfinder EXperiment telescope (APEX; G\"usten $\etal$ 2006) at the
Llano de Chajnantor high-elevation site in the Chilean Andes. The
first series of observations was carried out with the Carbon
Heterodyne Array of the MPIfR (CHAMP+) receiver in guaranteed
observing time between 2008 and 2012 (projects X-081.F-1002-2008,
E-085.B-0094B-2010, E-088.B-0075A.2011, and X-089.F-0007-2012). The
second series of observations was carried out with the Swedish-ESO PI
instrument for APEX (SEPIA) receiver in 2019 guaranteed observing time
and in 2021 regular European southern observatory (ESO) time (projects
E-0104.B-0034A-2019 and E-0108.C-0586A.2021).  At the observing
frequencies of 661.067 GHz ($\thirco$(6-5)), 691.473 GHz ($\co$(6-5))
and 806.651/809.350 GHz ($\co$(7-6)/$\ci$(2-1)) the APEX full-width
half maximum (FWHM) beam sizes are $9.3"$, $8.9"$ and $7.7"$ according
to the on-line data sheets. Calibration scans on Jupiter and Mars
yield efficiencies needed to transform antenna temperatures $T_{A}$
into main beam temperatures $T_{mb}$ of effectively $\eta_{mb}$ =
0.48, 0.52, and 0.48 with uncertainties of 0.02\footnote{
  www.mpifr-bonn.mpg.de/4482182/champ$\_$efficiencies$\_$16-09-14.pdf,
  www.apex-telescope.org/telescope/efficiency/}.  The conversion
factor $S/T_{mb}$ of flux density $S$ to brightness temperature
$T_{mb)}$ is about 60 Jy/K.
 
\subsection{CHAMP+ observations}

CHAMP+ is a dual-band 2 $\times$ 7 element heterodyne array developed
by the Max Planck Institute f\"ur Radioastronomie (MPIfR) in Bonn (D),
the Stichting RuimteOnderzoek Nederland (SRON) in Groningen (NL), and
the Jet Propulsion Laboratory (JPL) in Pasadena (USA). It is a
principal investigator (PI) instrument operated for the APEX community
as a collaborative effort with MPIfR (Kasemann $\etal$ 2006, G\"usten
$\etal$ 2008). The array can be operated simultaneously in ALMA
(Atacama large millimeter/submillimeter array) bands 9 and 10, and we
used this property to obtain carbon $J$=2-1 $\ci$ (rest frequency 809
GHz) and $J$=7-6 $\co$ (rest frequency 806 GHz) measurements
simultaneously with the Band 9 $J$=6-5 $\co$ and $\thirco$ line
measurements. Both sub-arrays have closely spaced pixels in a
hexagonal arrangement providing data sampling with half-beam spacing
in scanning mode. The backend is an autocorrelator array with a total
bandwidth of 32 GHz and 32768 spectral channels, subdivided into 32 IF
bands of 1 GHz and 1024 channels each.  We used position-switching
with a throw of $900"$, well clear of the galaxy main
bodies. On-the-fly maps were obtained for all sources, mostly with
$50"\,\times50"$ field-of-views. For the purposes of this paper, we
extracted single emission profiles by spatially binning all emission
within an area corresponding to the desired resolution. The Band 9
data were obtained with sky conditions varying from good (total system
temperature including sky $T_{sys}$ = 840 K) to just acceptable
($T_{sys}$ = 1675 K). In Band 10, total system temperatures varied
from 2400 to 4000 K. The calibration is estimated to be accurate to
$\leq30\%$. This error is almost entirely governed by baseline
uncertainties. The emission from the observed galaxies typically
occupies about half of the 1200 $\kms$ window covered by the backend
and leaves limited room for accurate baseline definition. The baseline
errors are too large to derive reliable fluxes for the weak $\thirco$
emission.

\subsection{SEPIA observations}

The (SEPIA) is a single-pixela heterodyne receiver with a cryostat
accommodating three ALMA-like receiver cartridges (Belitsky $\etal$
2018), provided by the group for advanced receiver development (GARD)
at the Onsala space observatory (S). We used the SEPIA660 cartridge
(Baryshev $\etal$ 2015), which is a dual polarization 2SB receiver
installed and commissioned by the Groningen NOVA group (NL) during the
second half of 2018.  The SEPIA660 receiver covers the window between
597 GHz and 725 GHz. It has two IF outputs per polarization, USB and
LSB, each covering 4-12 GHz, adding up a total of 32 GHz instantaneous
IF bandwidth. The central frequencies of the two side-bands are
separated by 16 GHz.  The backend was an FFT spectrometer with a
spectral resolution of about 61 kHz (26 m/s) with 65536 channels per
every 4 GHz.  For NGC~253 and NGC~4945 we obtained a five-point cross
on the central position in $\co$; all other observations were single
pointings.  The sky conditions mostly varied from very good ($T_{sys}$
= 500 - 700 K) to good ($T_{sys}$ = 700 - 1100 K). Throughout the
observations, the baselines were quite stable and the (much) wider
velocity coverage allowed good baseline definition and subtraction.

\subsection{Additional observations}

In the following discussion, we also include the mid-$J$ $\thirco$
observations of bright galaxies that already exist in the literature.
These concern spaceborne and ground-based detections of NGC~253
(Hailey-Dunsheath $\etal$ 2008, P\'erez-Beaupuits $\etal$ 2018),
IC~342 (Rigopoulou $\etal$ 2013), NGC~3034 (M~82; Loenen $\etal$ 2010,
Panuzzo $\etal$ 2010, Kamenetzky $\etal$ 2012), NGC~5128 (Centaurus~A;
Israel $\etal$ 2014) and NGC~4945 (Bellochi $\etal$ 2020).

\section{Results and analysis}

\begin{table}
\begin{center}
{\small %
\caption[]{\label{beamtable} Beam-dependent $\co$(6-5) intensities.}
\begin{tabular}{lrrrrccc}
\noalign{\smallskip}
\hline
\noalign{\smallskip}
  & \multicolumn{4}{c}{Intensity $I$(CO)=$\int${$T_{mb}$d$v$}} & \multicolumn{2}{c}{Fit Parameter$^a$} & Ratio$^b$ \\
              &$9"$ &$22"$&$33"$&$43"$     &  $a$   & $b$ & (6-5)/(1-0) \\
              & \multicolumn{4}{c}{($\kkms$)}      &        & & \\
 (1)     &(2)    & (3) & (4) & (5)            & (6)    & (7)  & (8) \\
\noalign{\smallskip}
\hline
\noalign{\smallskip}
N253$^c$ & 1049 & 459 & 255 & 237 & -1.00 & 3.97 & 0.45$\pm$0.05 \\
N660     &   48 &  20 &  14 &   8 & -1.08 & 2.94 & 0.13$\pm$0.02 \\
N1068    &   96 &  59 &  35 &  25 & -0.85 & 2.83 & 0.35$\pm$0.05 \\
N1097    &   49 &  20 &  12 &  10 & -0.81 & 2.32 & 0.15$\pm$0.03 \\
N1365    &   65 &  36 &  26 &  19 & -0.76 & 2.56 & 0.14$\pm$0.02 \\
N1808    &  112 &  38 &  26 &  19 & -1.13 & 3.12 & 0.28$\pm$0.05 \\
N3034$^d$&  313 & 177 & 147 & 131 & -0.56 & 3.02 & 0.26$\pm$0.03 \\
N4945$^e$&  724 & ... & 130 & ... & -0.74 & ...  & 0.20$\pm$0.04 \\
N5236    &   68 & ... & ... &  10 & -1.22 & ...  & 0.12$\pm$0.03 \\
N6240    &   55 & ... & ... &  10 & -1.09 & ..   & 0.30$\pm$0.1  \\  
  \noalign{\smallskip}     
\hline
\end{tabular}
}%
\end{center} 
Notes: a. slope $a$ and intercept $b$ a resulting from fitting of
observed intensities as a function of beamsize: log(intensity) = $a$
log(beam) + $b$.  b. $J$=1-0 data from Paper I; c. $40"$ from
$Herschel$-HIFI (P\'erez-Beaupuits $\etal$ 2018); d . Convolved $JCMT$
data (Seaquist $\etal$ 2006); $33"$ $Herschel$-HIFI from (Loenen
$\etal$ 2010); e. $33"$ from $Herschel$-HIFI data (Bellochi $\etal$
2020).
\end{table}

\begin{table}
\begin{center}
{\small %
\caption[]{\label{isotop} APEX $\co/\thirco$ isotopologue intensity ratios.}
\begin{tabular}{lcccc}
\noalign{\smallskip}     
\hline
\noalign{\smallskip} 
Name  & $J$=6-5$^a$ &$J$=3-2$^b$ & $J$=2-1$^c$ &$J$=1-0$^c$\\
      & ($9"$)      &  (14$"$)   &  ($11"$)  &  ($22"$)  \\
(1)   & (2)         &    (3)     &    (4)    &     (5)   \\
\noalign{\smallskip}      
\hline
\noalign{\smallskip}      
N253  &13.3$\pm$0.2$^d$&11.7$\pm$1.2& 10.7$\pm$1.1 & 12.7$\pm$1.3 \\
N613  & 15.1$\pm$4.4 & 11.3$\pm$2.1 & 10.5$\pm$2.1 & 11.5$\pm$1.8 \\
N660  & 12.7$\pm$3.3 & 12.4$\pm$1.9 & 19.8$\pm$2.4 & 14.0$\pm$1.4 \\
N1068 & 21.7$\pm$3.2 & 15.2$\pm$2.2 & 12.8$\pm$1.4 & 11.8$\pm$1.7 \\
N1097 & 11.0$\pm$3.7&14.6$\pm$2.3$^e$& 14.4$\pm$2.2 & 10.5$\pm$1.6 \\
N1365 & 13.6$\pm$2.0 & 12.2$\pm$1.3 & 11.7$\pm$1.2 & 11.1$\pm$15 \\
N1808 & 24.1$\pm$3.3 & 17.1$\pm$1.7 & 12.9$\pm$1.4 & 16.5$\pm$1.2 \\
N2559 & 13.7$\pm$2.7 & 22.4$\pm$4.3 & 10.4$\pm$2.1 &  9.9$\pm$1.9 \\
N4945 & 12.1$\pm$0.6 &7.5$\pm0.9$$^f$&7.7$\pm$0.9$^g$&13.2$\pm$1.3 \\
N5236 & 21.2$\pm$5.3 & 10.0$\pm$1.2 &  8.3$\pm$1.3 & 13.6$\pm$1.1 \\
N6240 & 24.8$\pm$6.8 & 26.5$\pm$6.6 & 38$\pm$9.4   & 29$\pm$8   \\ 
\noalign{\smallskip}
 \hline
\end{tabular}
}%
\end{center} 
Notes: a. APEX, This Paper; b. JCMT, Paper I; c. IRAM,
Paper I; d. CSO ($11"$), Hailey-Dunsheath $\etal$ (2008);
e. APEX, $18"$ Pi\~nol-Ferrer $\etal$ (2011); f. APEX ($18"$)
Bellochi $\etal$ (2020); g. $SMA$ ($10"$) Chou $\etal$ (2007);
\end{table}

\begin{table}
\begin{center}
{\small %
\caption[]{\label{litiso}$Herschel$ $\co/\thirco$ isotopologue intensity ratios.}
\begin{tabular}{lcccc}
\noalign{\smallskip}     
\hline
\noalign{\smallskip} 
Name      & $J$=5-4     & $J$=6-5   & $J$=7-6   & $J$=8-7   \\ 
          & ($35"$)     &($33"$)    &  ($36"$)  &  ($40"$)  \\ 
(1)       & (2)         &    (3)    &    (4)    &     (5)   \\
\noalign{\smallskip}      
\hline
\noalign{\smallskip}      
N253$^a$  & 18$\pm$1    & 25$\pm$8  & 29$\pm$11 & 30$\pm$12 \\ 
IC342$^b$ & 11$\pm$3    & 26$\pm$8  &    ...    &   ...     \\ 
N3034$^c$ & 19$\pm$4    & 24$\pm$2  & 30$\pm$6  & 34$\pm$10 \\ 
N4945$^d$ & 14$\pm$3    & 22$\pm$2  & 18$\pm$2  & ...       \\ 
Cen A$^e$ & 19$\pm$2    &  ...      &  ...      & ...       \\ 
\noalign{\smallskip}
\hline
\end{tabular}
}%
\end{center} 
Notes: 
a. SPIRE, P\'erez-Beaupuits $\etal$ (2018);
b. SPIRE, Rigopoulou $\etal$ (2013);
c. HIFI, Loenen  $\etal$, (2010); SPIRE, Panuzzo $\etal$ (2010),
Kamenetzky $\etal$ (2012);
d. HIFI, Bellochi $\etal$ (2020); SPIRE, Bellochi private communication (2021);
e. HIFI, Israel $\etal$ (2014).
\end{table}

\begin{figure*}
  \begin{center}
\begin{minipage}[]{18cm} 
  \begin{minipage}[]{5.06cm} 
    \resizebox{5.85cm}{!}{\rotatebox{0}{\includegraphics*{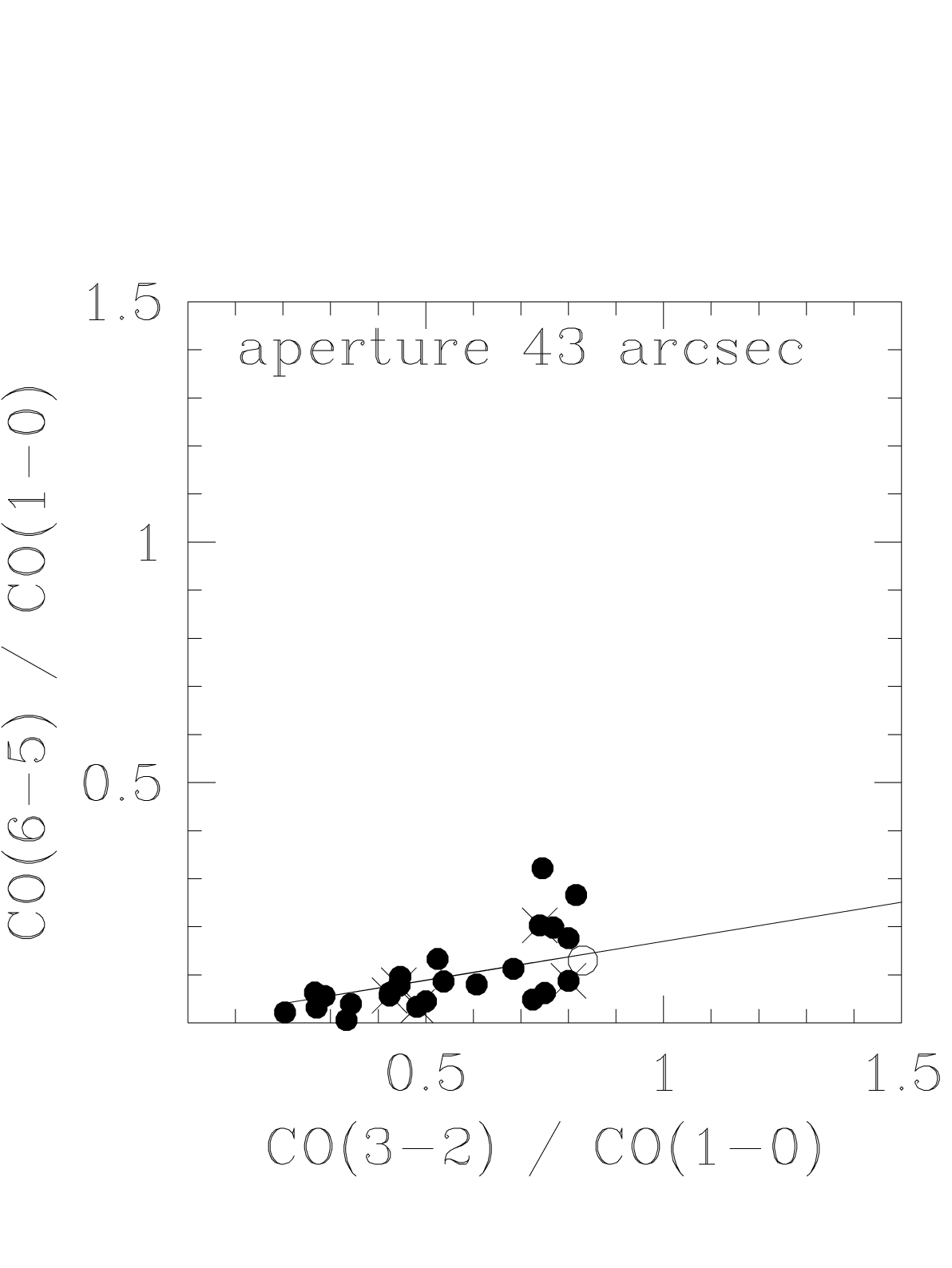}}}
  \end{minipage}
\hfill
  \begin{minipage}[]{5.06cm} 
    \resizebox{5.85cm}{!}{\rotatebox{0}{\includegraphics*{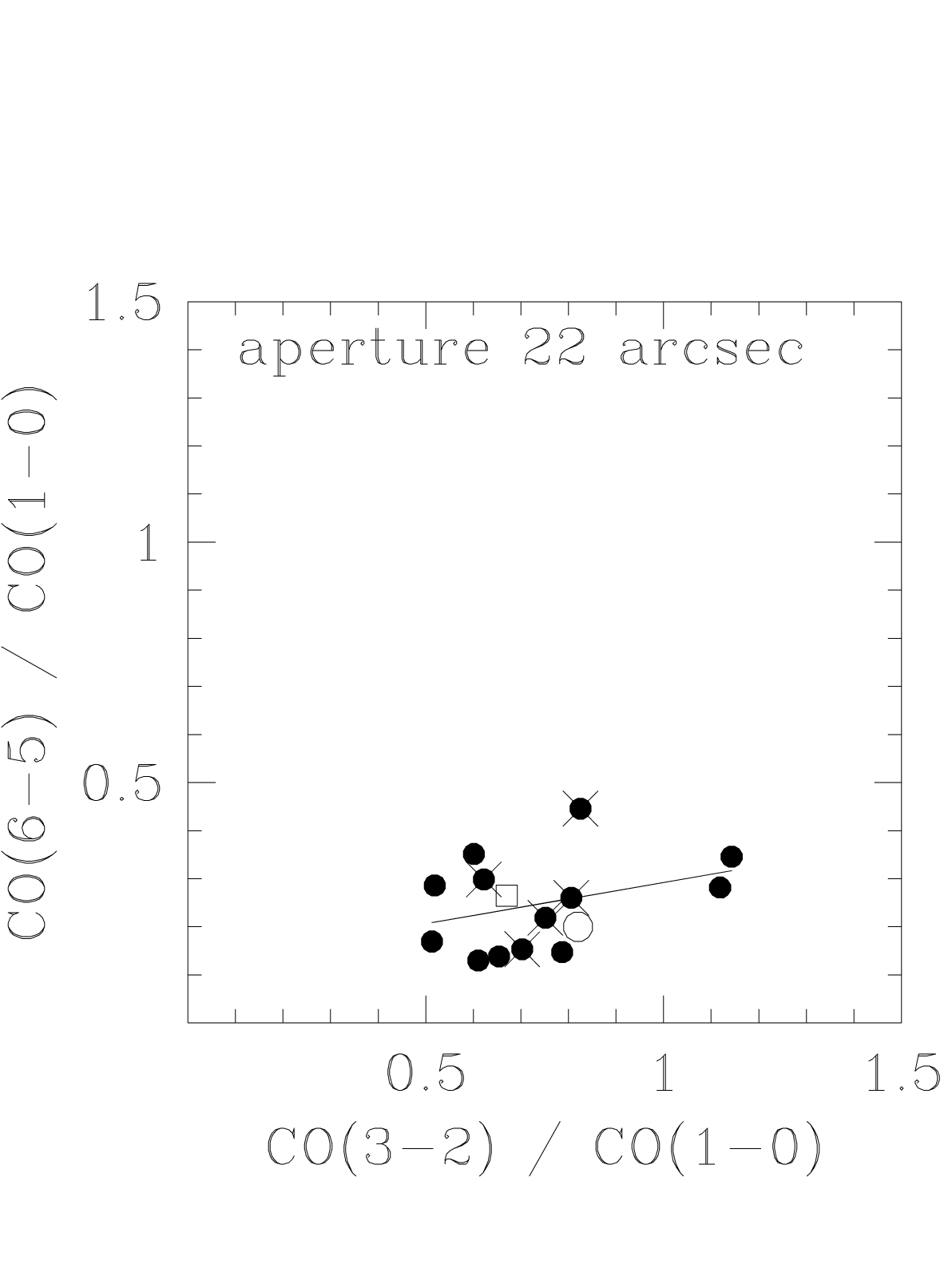}}}
  \end{minipage}
\hfill
  \begin{minipage}[]{5.06cm} 
    \resizebox{5.85cm}{!}{\rotatebox{0}{\includegraphics*{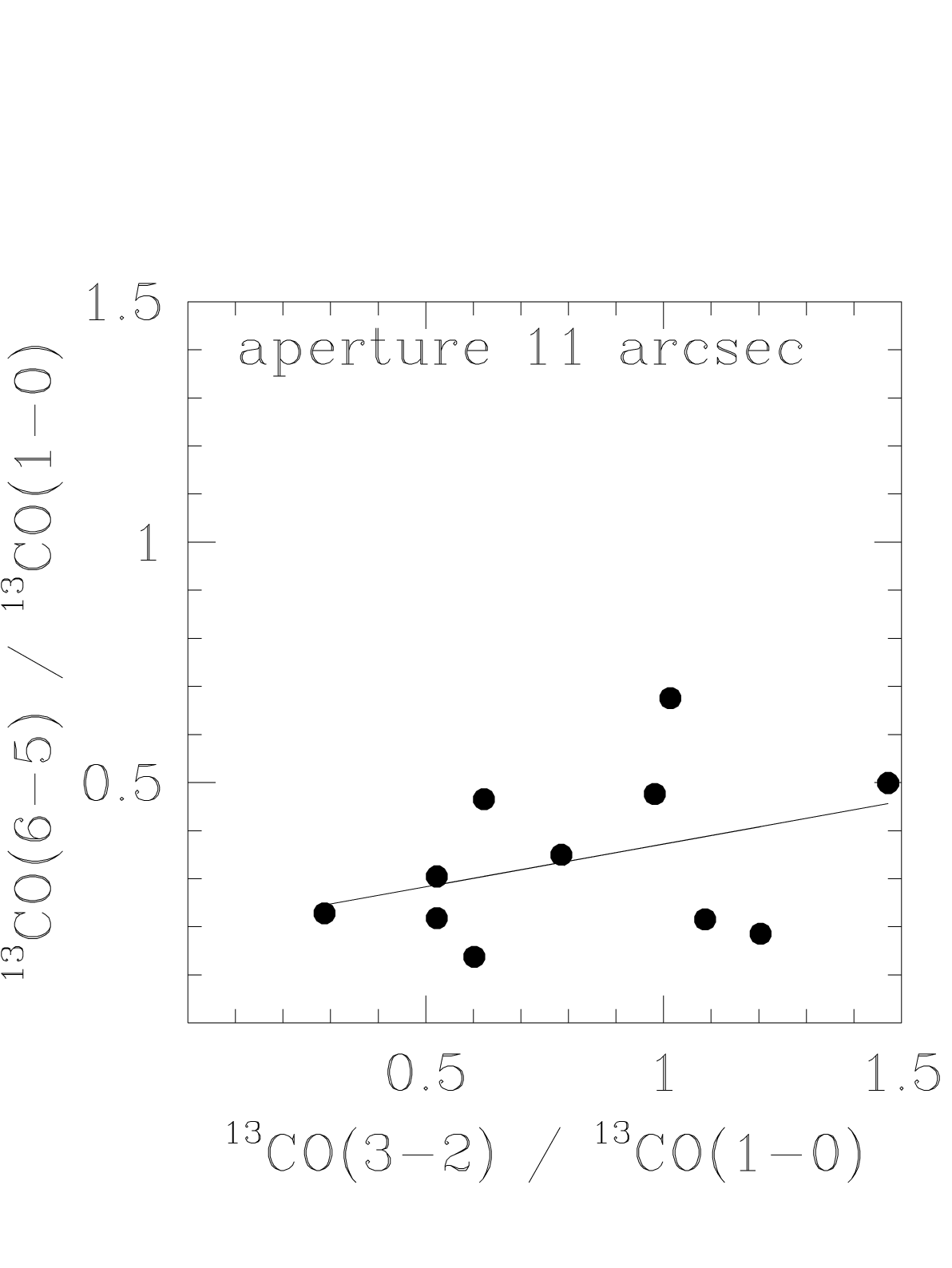}}}
  \end{minipage}
\end{minipage}
\begin{minipage}[]{18cm}
\vspace{-2cm}
  \begin{minipage}[]{5.06cm} 
    \resizebox{5.85cm}{!}{\rotatebox{0}{\includegraphics*{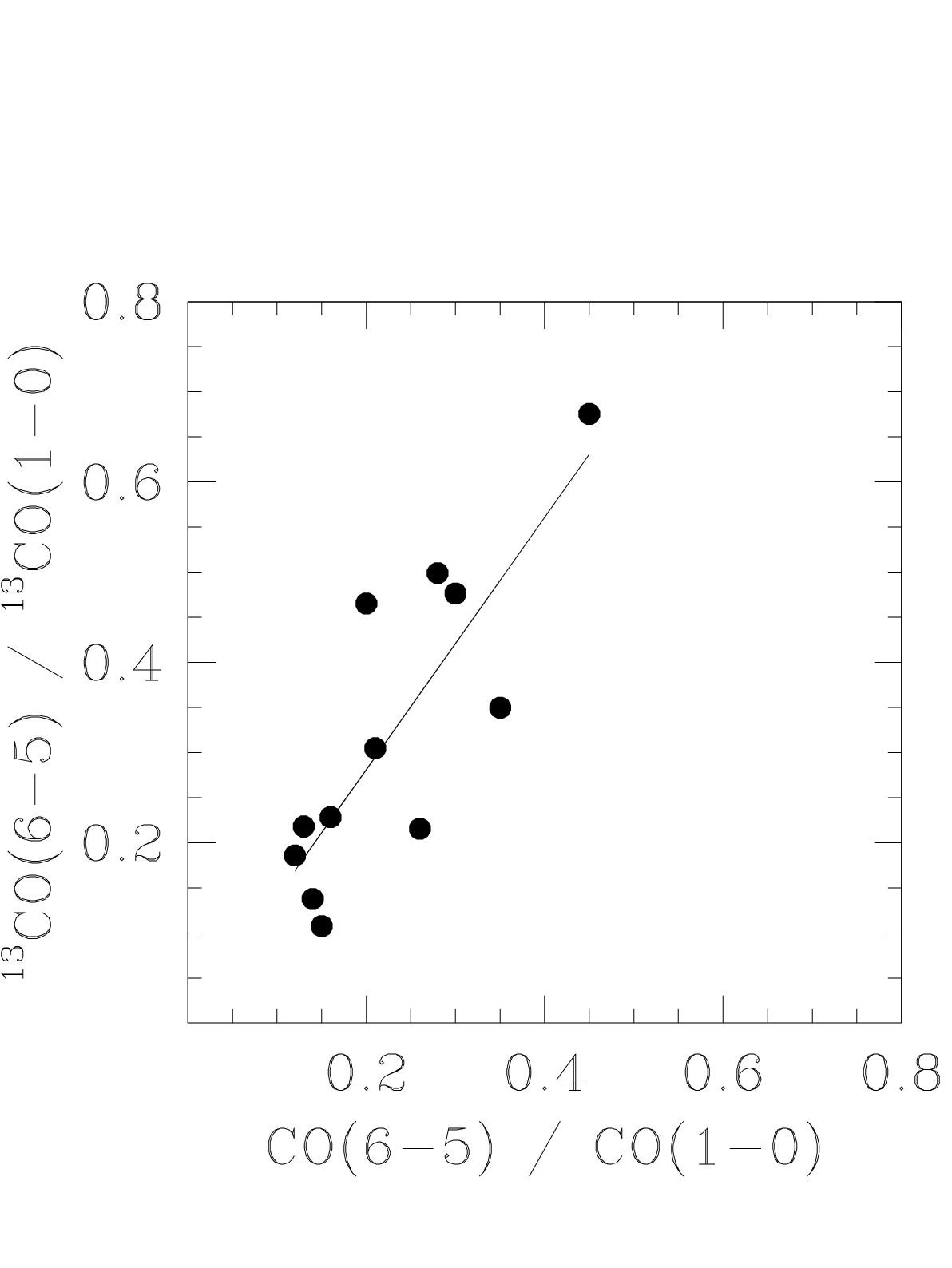}}}
  \end{minipage}
\hfill
  \begin{minipage}[]{5.06cm} 
    \resizebox{5.85cm}{!}{\rotatebox{0}{\includegraphics*{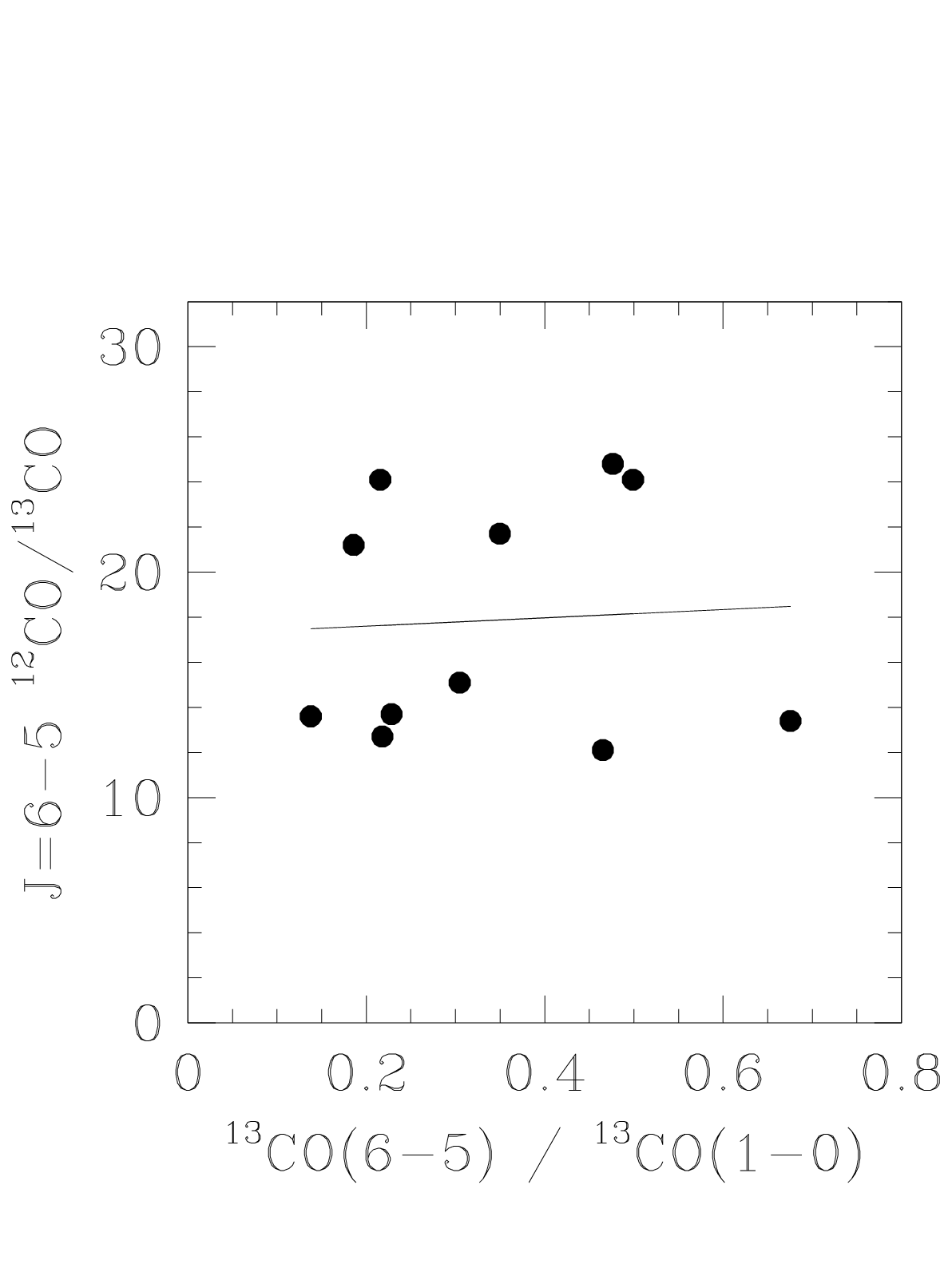}}}
  \end{minipage}
\hfill
  \begin{minipage}[]{5.06cm} 
    \resizebox{5.85cm}{!}{\rotatebox{0}{\includegraphics*{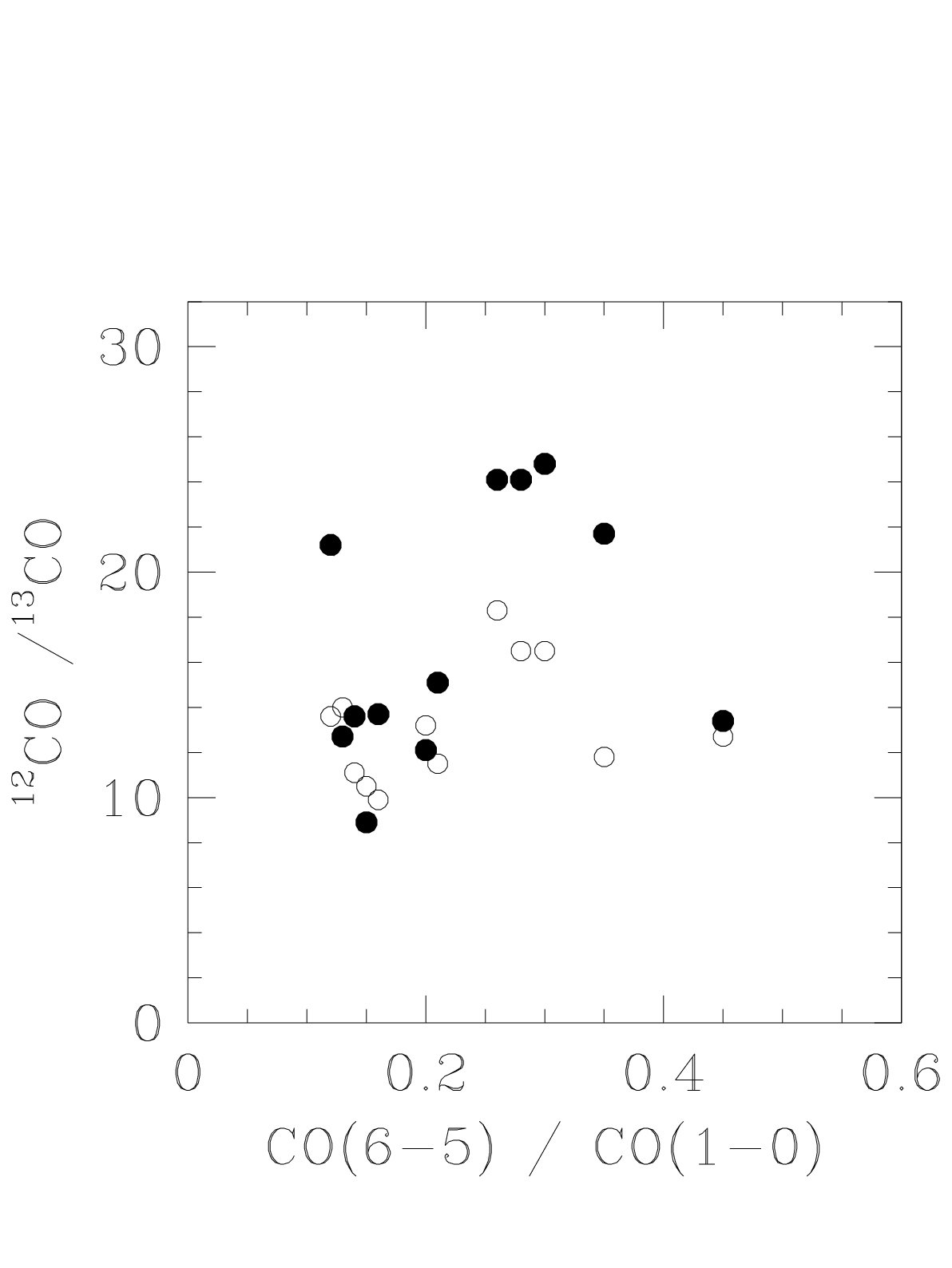}}}
  \end{minipage}
\end{minipage}
\caption[] {Comparison of observed line ratios. Straight lines mark
  least-squares linear fits to the data displayed. Top row: comparison
  of $J$=6-5 and $J$=3-2 line intensities normalized by $J$=1-0
  intensity. Left: $\co$ ratios in $43"$ apertures. Open circle marks
  the Inner Galaxy including Galactic Nucleus. Center: $\co$ ratios in
  $22"$ apertures. Open circle marks the Galactic Center; open square
  marks the starburst dwarf galaxy He2-10. In both panels, the
  relatively nearby galaxies at distances $D\leq6.5$ Mpc are marked by
  a cross. Right: $\thirco$ ratios in extrapolated $11"$
  apertures. Bottom row: comparison of isotopologue intensities 
    in (extrapolated) $11"$ apertures.  Left: comparison of the
  $\thirco$ and $\co$ $J$=6-5/$J$=1-0 intensity ratios. Center:
  comparison of the $J$=6-5 isotopologue intensity ratios and the
  $\thirco$ $J$=6-5/$J$=1-0 intensity ratios. The outlier NGC~6240 is
  not included.  Right: isotopologue intensity ratios as a function of
  $\co$(6-5)/$\co$(1-0) intensity ratios. The $J$=1-0 isotopological
  ratios are assumed to be identical in $11"$ and $22"$ apertures.
  Filled circles mark isotopological intensity ratios in the $J$=6-5
  transition, open circles those in the $J$=1-0 transition. }
\label{corat}
\end{center}
\end{figure*}

\begin{figure*}
  \begin{center}
\begin{minipage}[]{18cm} 
  \begin{minipage}[]{5.06cm} 
    \resizebox{5.86cm}{!}{\rotatebox{0}{\includegraphics*{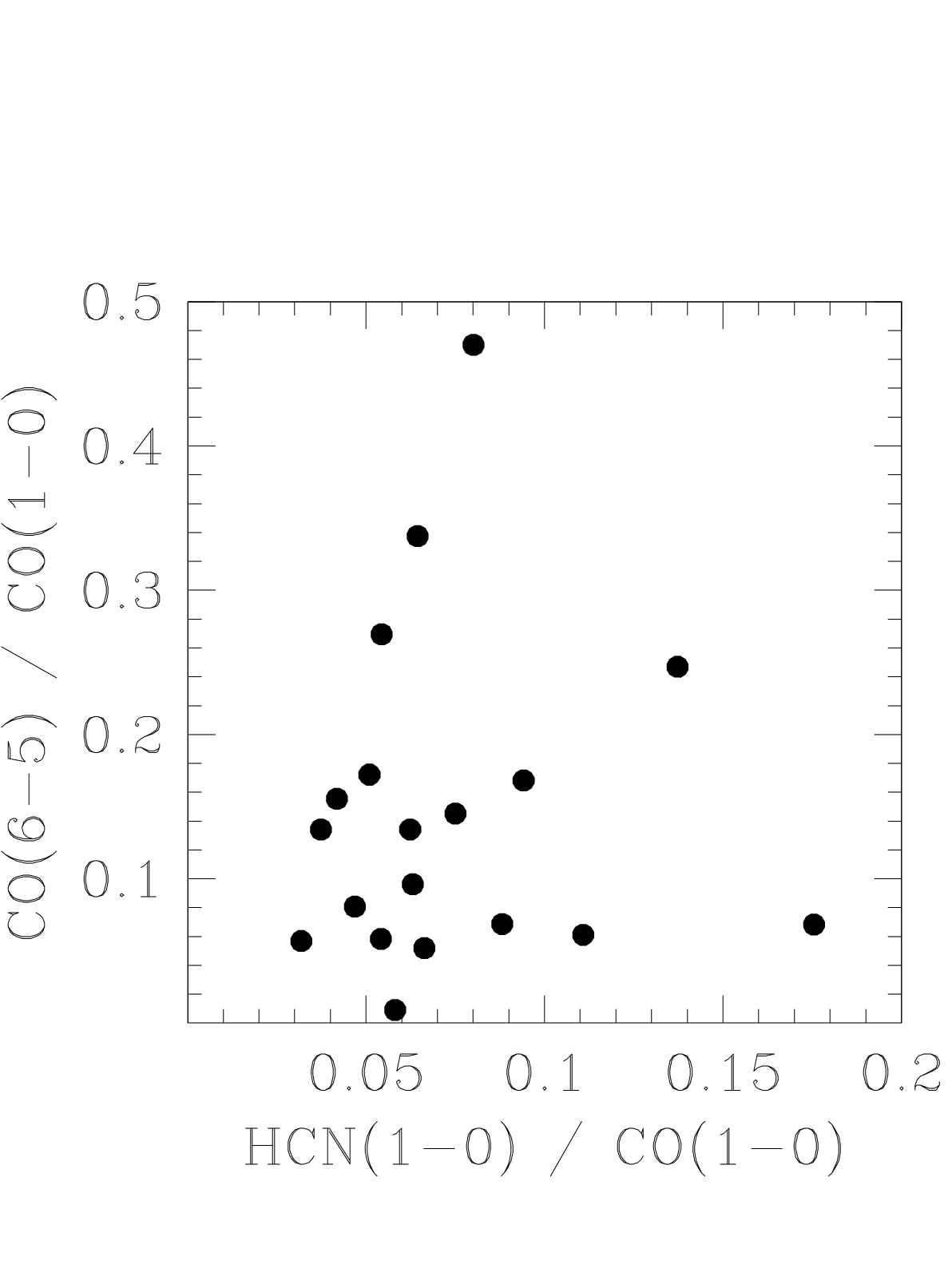}}}
  \end{minipage}
\hfill
  \begin{minipage}[]{5.06cm} 
    \resizebox{5.86cm}{!}{\rotatebox{0}{\includegraphics*{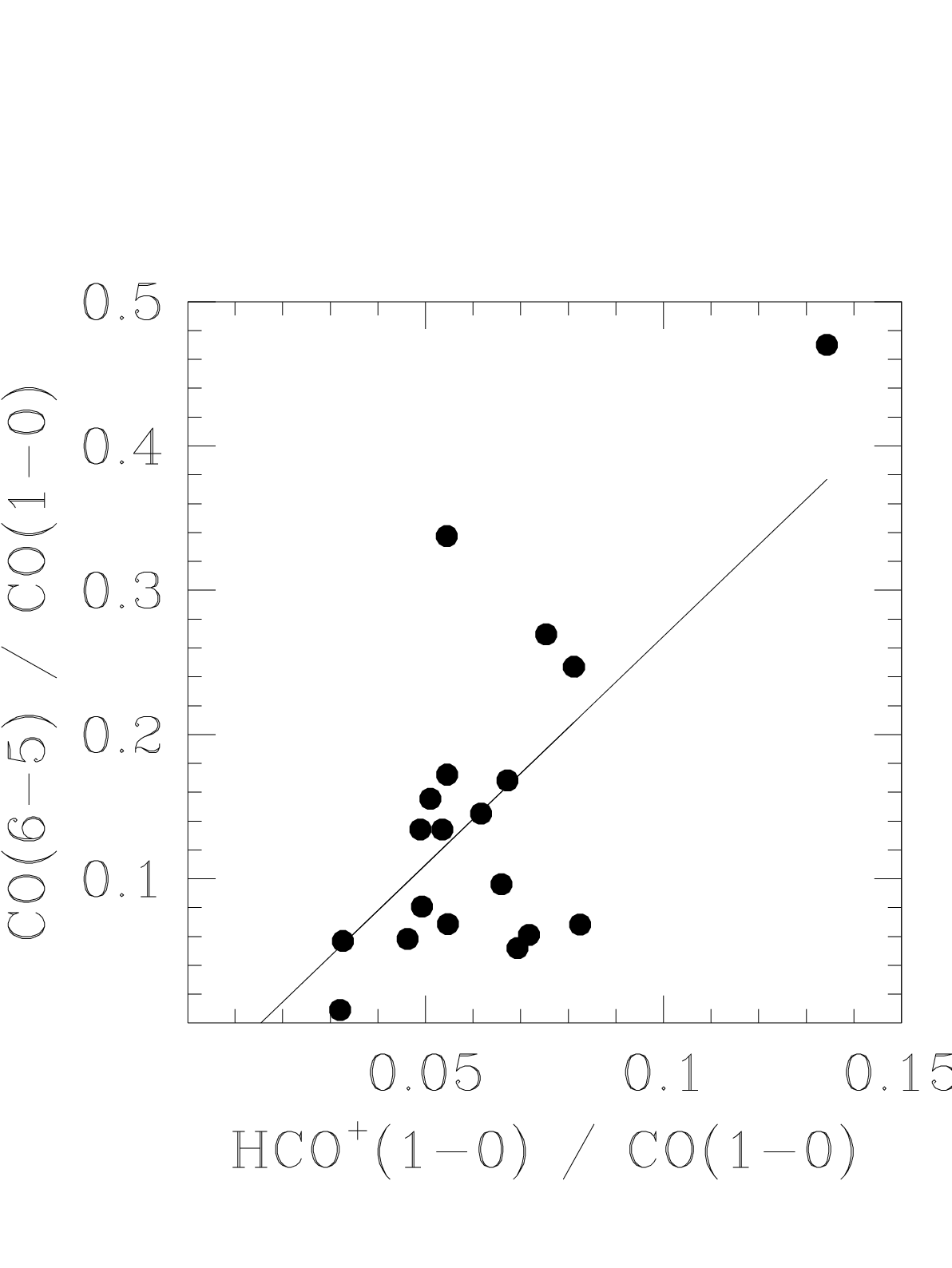}}}
  \end{minipage}
\hfill
  \begin{minipage}[]{5.06cm} 
    \resizebox{5.86cm}{!}{\rotatebox{0}{\includegraphics*{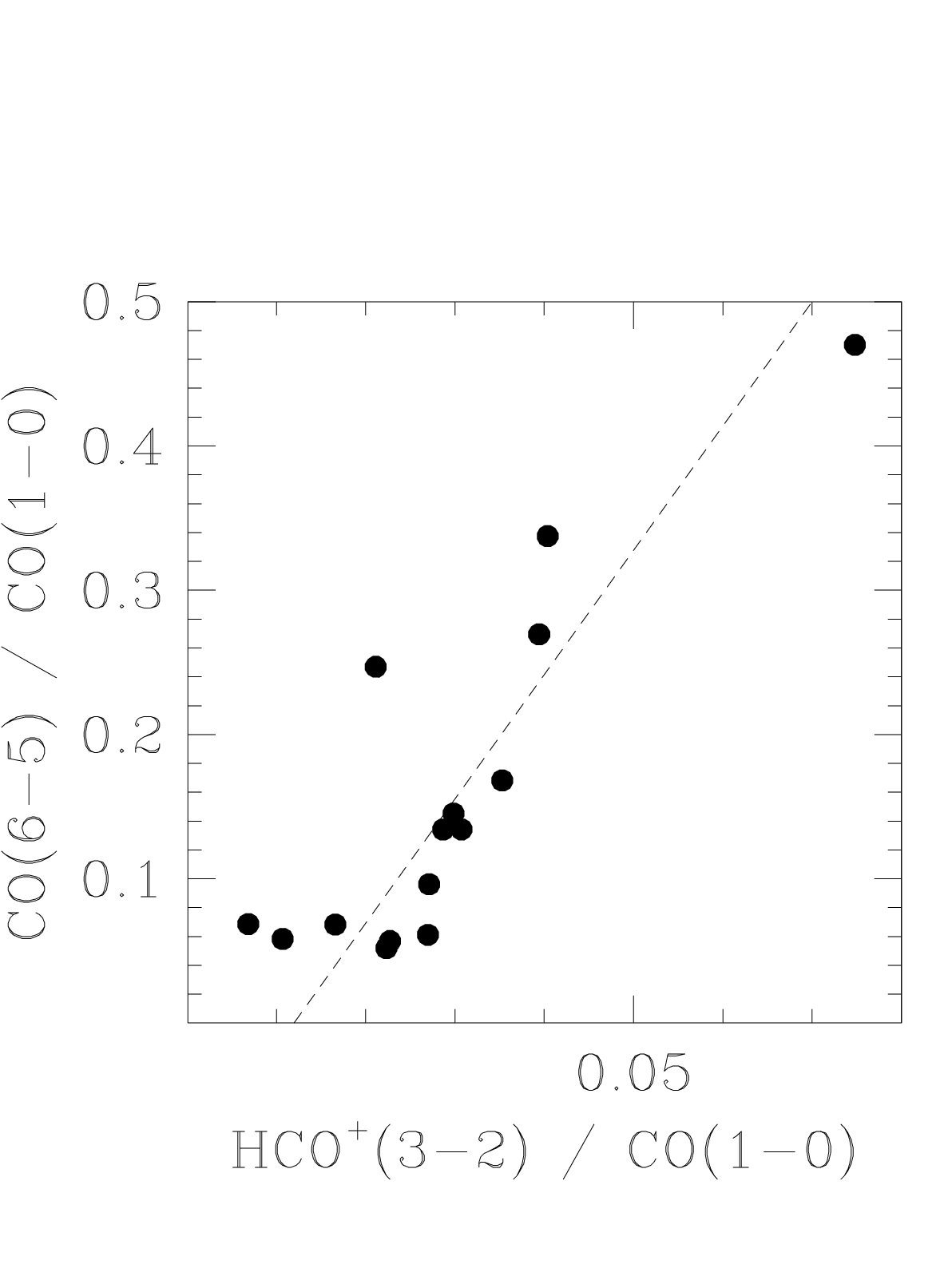}}}
  \end{minipage}
\end{minipage}
\begin{minipage}[]{18cm}
    \vspace {-2cm}
  \begin{minipage}[]{5.06cm} 
    \resizebox{5.86cm}{!}{\rotatebox{0}{\includegraphics*{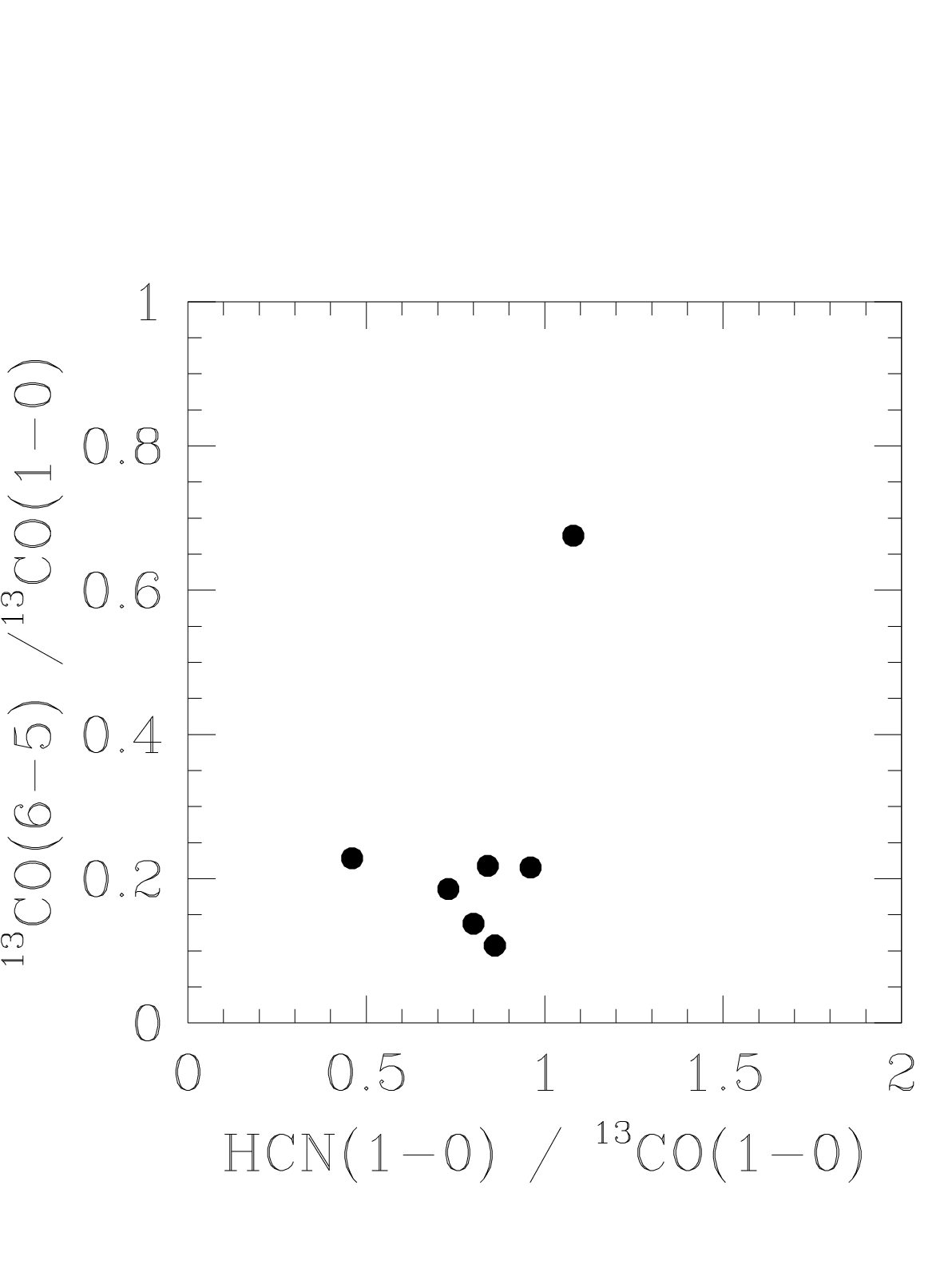}}}
  \end{minipage}
\hfill
  \begin{minipage}[]{5.06cm} 
    \resizebox{5.86cm}{!}{\rotatebox{0}{\includegraphics*{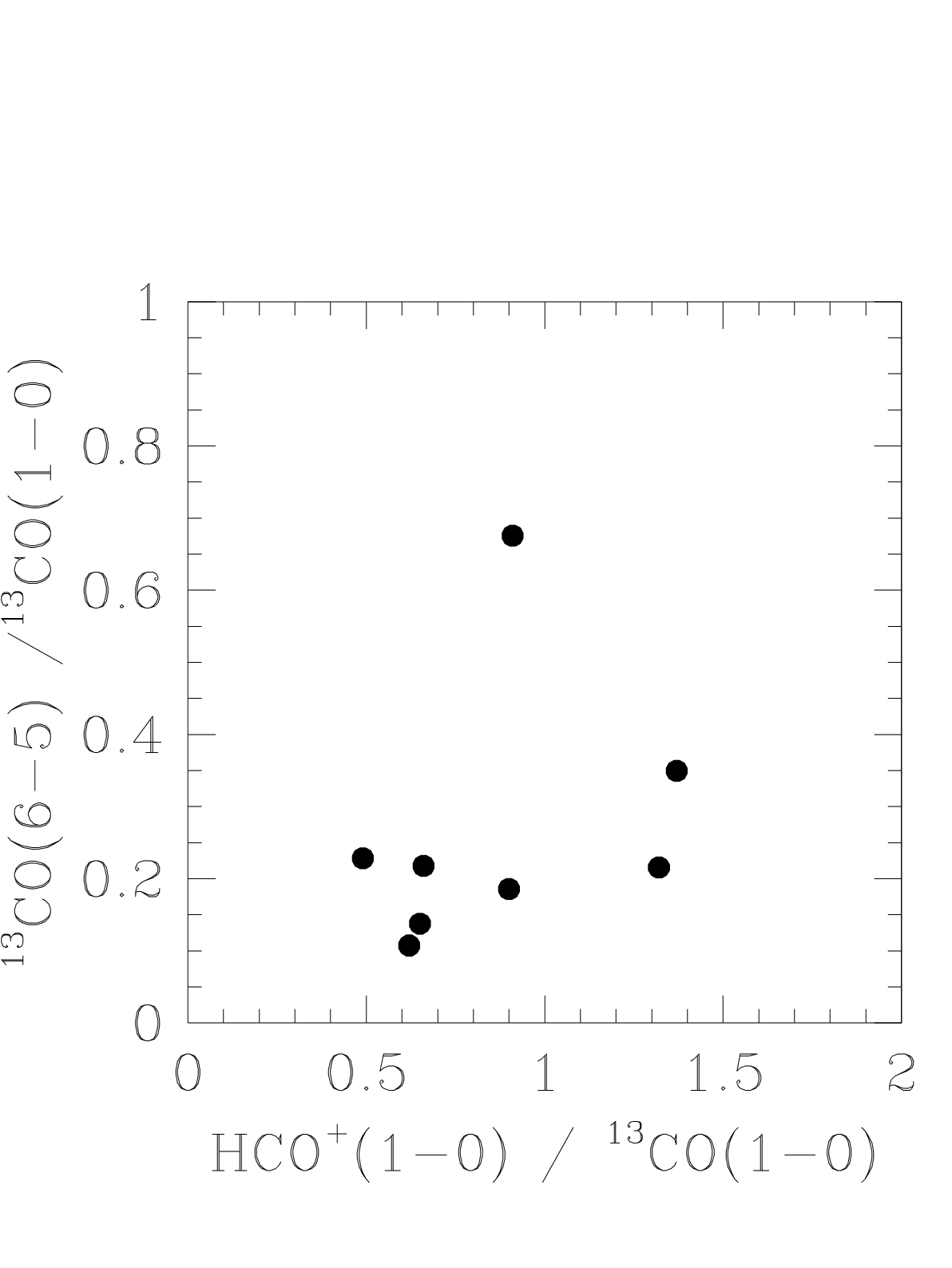}}}
  \end{minipage}
\hfill
  \begin{minipage}[]{5.06cm} 
    \resizebox{5.86cm}{!}{\rotatebox{0}{\includegraphics*{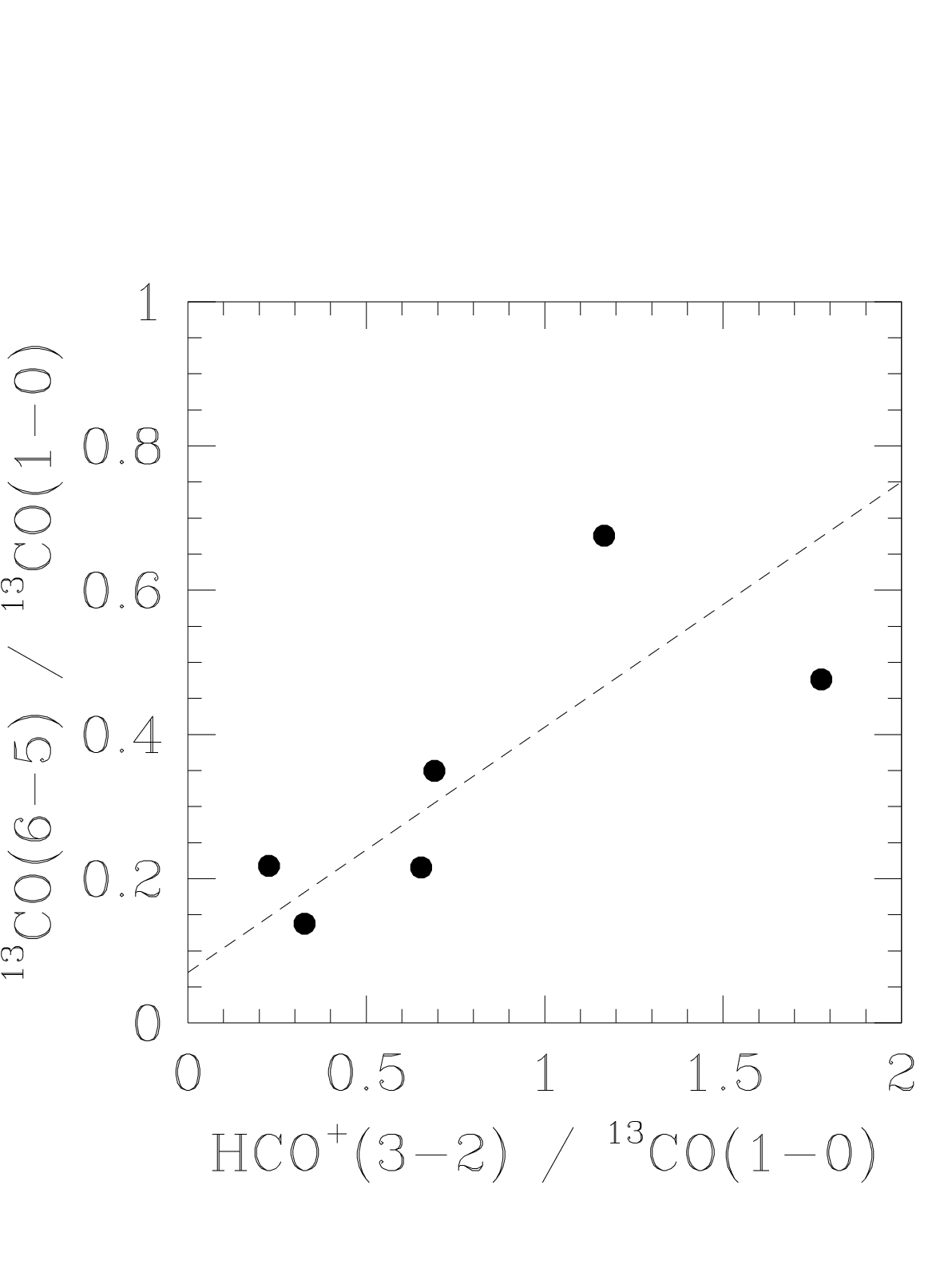}}}
  \end{minipage}
\end{minipage}
\caption[] {Top: $\co$ (6-5)/(1-0) intensity ratios as a function of
  HCN(1-0), HCO$^+$ and HCO$^+$(3-2) line intensities relative to
  $\co$(1-0). All ratios refer to a resolution of $22"$, except the
  $\thirco$ (6-5)/(1-0) ratios that refer to extrapolated
  $11"$ apertures (cf. section 4.3). The central and rightmost panels
  include linear fits to the data. Bottom: the same as the top row,
  but with $\thirco$ lines substituted for $\co$ lines. The dashed
  line in the rightmost panel is an eyeball fit to the limited data}
\label{molrat}
\end{center}
\end{figure*}

\subsection{Extracted $\co$ line intensities}

The velocity-integrated and peak $J$=6-5 $\co$ and $\thirco$ line
intensities measured with the CHAMP+ and SEPIA660 receivers are listed
in Table\,\ref{results}.  The table also lists the $J$=7-6 $\co$ and
$J$=2-1 $\ci$ fluxes measured with the CHAMP+ receiver in parallel
with the $J$=6-5 observations.  To facilitate comparison, we have
binned these higher frequency data to the somewhat lower spatial
resolution of the $J$=6-5 measurements.  The errors listed are
stochastic errors; they do not include systematic errors due to
calibration etc.  For NGC~1365 and NGC~1808 $\co$ intensities were
available from both the CHAMP+ and the SEPIA660 receiver on APEX.
Comparison shows a good agreement between the results obtained with
the different receivers on different occasions. In spite of the
different observing techniques, there is likewise good agreement, for
the three galaxies observed with $Herschel$/SPIRE that were also
mapped with APEX/CHAMP+. In comparable apertures, the CHAMP+ fluxes
are 20 per cent higher for both NGC~1068 and NGC~1097 and 20 per cent
lower for NGC~1365. The galaxies NGC~660 and NGC~1808 were not
observed with SPIRE.

Five of the sample galaxies are relatively close to our Galaxy with
distances $D$=3.4-4.5 Mpc (Table\,\ref{sample}). At microwave
frequencies, they are among the brightest galaxies in the sky. Three
of them (NGC~253, M~82, and NGC~4945) accordingly have $\co$(6-5)
intensites an order of magnitude higher than the other galaxies in the
survey, including those at the same distance (IC~342 and M~83). The
remaining eight galaxies have $\co$ intensities similar to the latter
although they are at three to five times greater distance, thirty
times greater in the case of the ultra-luminous galaxy NGC~6240. A
similar pattern applies to the considerably weaker $\thirco$(6-5)
fluxes that required integration times up to several hours.

In ten of the thirteen galaxies of the sample, $\co$(6-5) intensities
were measured at different spatial resolutions. They either have
multiple pointed observations in the different SEPIA660, HIFI and
SPIRE apertures, or they were mapped, for instance with the CHAMP+
receiver, which allowed extraction of intensities binned to various
resolutions $\theta$. The results for individual galaxies are shown in
Fig.\,\ref{beamfig} together with fits to the data of the form
log($T_{mb}$d$v$) = a log($\theta$) + b.  In this formulation, $a$=0
corresponds to an extended source of constant surface brightness and
$a=-2$ corresponds to a point source. Intensities at the various
resolutions in the narrow range $-0.75\geq a\geq-1.2$ ($9"$ to $43"$)
are listed in Table\,\ref{beamtable} together with the fit
coefficients $a$ (slope) and $b$ (intercept).  The nearly edge-on
galaxy NGC~3034 (M~82) has a flat slope $a=-0.56$ caused by a relative
lack of $\co$(6-5) emission from the center (see Seaquist $\etal$
2006, Loenen $\etal$ 2010) but the remainder has an average slope
$a=-0.92$. A similar inverse proportionality between
intensity and aperture in the central regions of galaxies was also
found, with larger dispersion, for low-$J$ emission from the $\co$
molecule (Paper I) as well the HCN and HCO$^+$ molecules that trace
molecular gas with critical densities similar to those of the
$\co$(6-5) transition (Israel 2023, hereafter Paper II). Such behavior
is characteristic for the centrally peaked emission from galaxies
illustrated in Figure 3 of Paper I.  No dependence on distance is
apparent. The average $a$=-0.9 of the nearby galaxies NGC~253,
NGC~3034, NGC~4945, and NGC~5236 only samples the central 0.7-0.9
kpc. The intensities of the more distant galaxies (except NGC~6240)
sample the significantly larger inner regions with diameters of
2.6-4.5 kpc, yet have the same average $a$=-0.9.

\subsection{$J$=6-5 $\co/\thirco$ intensity ratios}

The APEX survey provides direct determination of the $J$=6-5
$^{12}$CO/$^{13}$CO intensity ratio at a resolution of $9"$ in the
eleven galaxies listed in Table\,\ref{isotop}.  The table includes the
corresponding $J$=3-2, $J$=2-1, and $J$=1-0 ratios in apertures
covering surface areas larger by factors of 2.4, 15, and 6.0,
respectively (Paper I). The $J$=1-0 to $J$=3-2 intensity ratios
are typically 10-15 in all galaxies except NGC~6240. In seven galaxies
the $J$=6-5 and $J$=1-0 ratios do not significantly differ. In the
other galaxies, the $J$=6-5 ratio exceeds the $J$=1-0 ratio by a
factor of up to two. Four galaxies have relatively high $J$=6-5 ratios
$\>20$. The intensity ratios of the LIRG NGC~6240 are the highest in
the sample. Papadopoulos $\etal$ (2014) suggest even higher ratios for
this galaxy. These and other literature ratios have, however, large
uncertainties because the very weak and broad $\thirco$ lines are
highly sensitive to baseline errors. Measurements with $Herschel$
provide additional $J$=5-4 to $J$=8-7 intensity ratios for five more
galaxies at the substantially lower resolutions of $33"$-$43"$ listed
in Table\,\ref{litiso}, thereby sampling surface areas twelve to
twenty times larger than covered by the APEX beam.  The
large-aperture $\co/\thirco$ intensity ratios in the $J_{upp}\geq6$
transitions in Table\,\ref{litiso} are two to three times higher.  The
$J$=5-4 intensity ratio is transitional, with an in-between average of
$\sim$16.  The large-area $Herschel$ $J$=6-5 ratios vary little,
ranging between 22 and 26, and are similar to the four highest $J$=6-5
ratios in the APEX sample. For NGC~253 and NGC~4945, $J$=6-5 ratios
are available in both apertures. In the fifteen times larger surface
area sampled by Herschel, the $J$=6-5 isotopologue ratio has increased
from the APEX value $\sim13$ to the almost twice higher value of
23. In these two galaxies, the $\thirco$ intensities are thus much
weaker beyond the inner 200 pc and even more centrally peaked than the
$\co$ intensities. At least half (six out of thirteen) of the
galaxies surveyed exhibits $J_{upp}\geq6$ intensity ratios of 20-30 in
either small or large apertures. By analogy, this could also be the
case for the other galaxies in Table\,\ref{isotop} lacking $Herschel$
$\thirco$ data.

We can take this a step further for NGC~253 and NGC~4945. After
subtraction of the contribution by the inner circumnuclear area
(APEX) to the larger central area ($Herschel$), the intensity ratio
in the residual $9"$-$33"$ zone of either galaxy becomes
$\co(6-5)/\thirco(6-5)\sim40$, suggesting low optical depths also for
$\co$. Together with ratios in excess of 20-30 in transitions with
$J_{upp}\geq6$ this is consistent with intrinsic [$\co$]/[$\thirco$]
isotopologue abundances of about 40 (cf. Tang $\etal$ 2019, Viti
$\etal$ 2020), as assumed in our earlier Papers I and II on
circumnuclear molecular gas.

\subsection{Further line ratio comparisons}

In these papers, we adopted a ``standard'' aperture of $22"$ for
intercomparison of the observed line intensities. For the galaxy
centers in this paper, we obtained such normalized $\co$(6-5)
intensities in a $22"$ aperture either by direct determination or
interpolation from Table\,\ref{beamtable}, or by extrapolating the
observed $9"$ intensities from Table\,\ref{results} with the average
slope $a$=-0.9 just determined (NGC~613, IC~342, and NGC~2559). These
are directly comparable to the $22"$ $J_{upp}\leq4$ line intensities
(Paper I) of the same galaxies. In Fig.\,\ref{corat} (top) we show
relations between $\co$ line ratios constructed from these data. We
also plotted transition intensity ratios in the $43"$ apertures of a
larger sample of twenty galaxies using the $\co$ intensities compiled
by Kamenetzky $\etal$ (2016) and Paper I. Nine galaxies are common to
both samples. For comparison we added the nearby starburst dwarf
galaxy He2-10 (data from Bayet $\etal$ 2006) as well as both the inner
Milky Way and the Galactic Center (data from Fixsen $\etal$ 1999)

As noted in section 4.2, comparison of the data in
Tables\,\ref{isotop} and \ref{litiso} suggests a strong
aperture-dependency of the $\co/\thirco$ ratio.  In order to limit the
extrapolation as much as possible, we have adopted for this case a
common aperture of $11"$. We assume identical distributions for the
$J$=6-5 emission in the $\thirco$ and $\co$ lines over this small
range.  If instead the $\thirco$(6-5) emission is point-like, its
normalized intensity is overestimated by $\sim15\%$. For the
normalization of the $\thirco$(1-0) intensity we assumed emission
aperture ratios identical to those in the $J$=2-1 transition (Paper
I). In a similar way we extrapolated the $\thirco$(3-2) data from
$14"$ to $11"$.

The $\thirco$(6-5) intensities are not so easily normalized to a $22"$
aperture because only a single point per galaxy is observed at a
resolution of $9.5"$.  In order to minimize aperture-dependent effects
we have limited the extrapolation to an aperture of $11"$ and we
assumed identical distributions for the emission in the $thirco$ and
$\co$ lines over this small range.  If instead the $\thirco$(6-5)
emission is point-like, its normalized intensity is overestimated by
$\sim15\%$. For the normalization of the $thirco$(1-0) intensity we
assumed instead emission aperture ratios identical to those in the
$J$=2-1 transition (Israel 2020). In a similar way we extrapolated the
$\thirco$(3-2) data over the small range of $14"$ to $11"$. In
Figure\,\ref{corat} (bottom) we show the relations of the
isotopologues to each other and to the $J$=6-5 and $J$=1-0
transitions.

\subsection{Excitation of the $\co$(6-5) gas}

Studies of the $\co$ rotational ladders of galaxy centers have been
published by several authors (Bayet $\etal$ 2006, Weiss $\etal$ 2007,
eve $\etal$ 2014, Rosenberg $\etal$ 2015, Mashian $\etal$ 2015, and
Kamenetzky $\etal$ 2016) to which we refer for further detail. In
these studies, the rotational ladders are primarily interpreted in
terms of excitation and the heating and cooling balance of the
gas. The SPIRE and PACS $\co$ ladders (Rosenberg $\etal$ 2015, Mashian
$\etal$ 2015) illustrate the great variety in overall shape. This
variety is already apparent in the line intensity ratios shown in the
top left and center panels of Fig.\ref{corat}. The excitation
represented by the (6-5)/(1-0) and the (3-2)/(1-0) $\co$ ratios varies
for the galaxies in our sample in a manner not related to galaxy
type. The excitation of the emission in these transitions increases
significantly with decreasing aperture size. The trend is continued
when $\thirco$ line intensity ratios observed at even higher
resolution are plotted (Fig.\,\ref{corat} top right). The two ratios
are weakly correlated (slope a=0.17$\pm$0.09). The systematic
displacements imply that the excitation of the central molecular gas
increases towards the galaxy nucleus.

The panels in the bottom row of Fig.\,\ref{corat} show the relations
of the isotopological $\co$ and $\thirco$ intensities to each other
and to the $J$=6-5/$J$=1-0 transitions ratios.  Although the LIRG
NGC~6240 was observed, we exclude it from most of the following
analysis as its extreme distance, surface area measured, and
luminosity class (see e.g.,  Greve $\etal$ 2009, Papadopoulos $\etal$
2014) set it too far apart from the other galaxies in the sample. For
the other galaxies, there is a clear relation between the $\thirco$
and $\co$ ladders (bottom left panel) but the intensity ratio of the
(6-5) and (1-0) transitions increases more rapidly for the optically
thin $\thirco$ than for the optically thick $\co$. On the other hand,
the $J$=6-5 $\co/\thirco$ isotopological intensity ratio is not
correlated with the intensity ratio of the $J$=6-5 and $J$=1-0
$\thirco$ or $\co$ transition that track the gas excitation nor is the
$J$=1-0 $\co/\thirco$ isotopological intensity ratio (bottom center
and bottom right panels).

The critical density of $\co$(6-5) is similar to that of HCN(1-0) and
falls in between those of HCO$^+$(1-0) and HCO$^+$(3-2), (see
Table\,\ref{crit}, so that a mutual comparison may be of
interest. Again, we normalize all line intensities by the $\co$(1-0)
intensity. Fig.\,\ref{molrat} explores the behavior of $J$=6-5 $\co$
and $\thirco$ lines as a function of the HCN(1-0), HCO$^+$(1-0) and
HCO$^+$(3-2) intensities (data from this paper and from Papers I and
II).  In all panels, the most extreme (6-5)/(3-2) $\co$ ratios belong
to NGC~6240 (high) and NGC~5055 (low). No correlation is apparent
between the HCN(1-0) and either $\co$(6-5) or $\thirco$(6-5) line
intensities (Fig.\,\ref{molrat}, leftmost panels) or HNC(1-0) (not
shown) despite their very similar critical densities.  There is,
perhaps, a correlation between the $\co$(6-5) and HCO$^+$(1-0) line
emission (top center panel) and, more convincingly in spite of the few
data points available, between HCO$^+$(3-2) and $\co$(6-5) and even
$\thirco$(6-5) (rightmost top and bottom panels).  This suggests that
HCO$^+$ is linked to the excitation of high-$J$ $\co$ and $\thirco$
(but see also Papadopoulos $etal$ 2010) and HCN is not. This is
consistent with the poor sensitivity of the HCN/CO intensity ratios to
both column density and fraction of dense gas noted by Priestley
$\etal$ (2024) in molecular cloud simulations and by Israel (2023) in
extragalactic multi-transition molecular line surveys.  Although the
heating and cooling of extragalactic molecular gas can be determined
from the observed $\co$ ladders, this is not so easily the case for
its physical parameters temperature, density, column density as these
are highly degenerate. Single-gas-phase models in general do not
adequately explain even relatively uncomplicated extragalactic $\co$
ladders and models with two or more distinct gas components are needed
(e.g.,  Greve $\etal$ 2014, Mashian $\etal$ 2015, Kamenetzky $\etal$
2016).  This need had already been reacognized in the analysis of
multiple low-$J$ transitions of optically thick $\co$ complemented by
optically thin $\thirco$ transitions (for instance Israel 2001, 2005;
Bayet $\etal$ 2006).

\section{Molecular gas physics revealed by $J$=6-5 CO lines}


\begin{table*}
\begin{center}
{\small %
\caption[]{\label{65model}Physical parameters from molecular gas modeling.}
\begin{tabular}{lcccccccccc}
\noalign{\smallskip}     
\hline
\noalign{\smallskip} 
Name     &\multicolumn{3}{c}{Gas parameters}     & Pressure  &\multicolumn{3}{c}{Fractional}         &\multicolumn{3}{c}{Model combined intensity ratios} \\
         & Density  & Temperature & Gradient     & parameter &\multicolumn{2}{c}{emission} &  mass   & $\co$    &\multicolumn{2}{c}{$\co/\thirco$} \\
         & $n(\h2)$ & $T_{\rm kin}$ &  $N_{\rm CO}$/d$v$ & $nT$ & $\co$(1-0) & $\co$(6-5) &      &(6-5)/(1-0)&$J$=1-0  & $J$=6-5\\
         &($\cc$)  &  (K)        & ($\cm2/\kms$) &($\cc$ K)  &        &            &                 &           &         &        \\
  (1)    &  (2)    &  (3)        &  (4)          & (5)       & (6)    & (7)        & (8)             &  (9)      & (10)    &  (11)   \\
\noalign{\smallskip}     
\hline
\noalign{\smallskip} 
NGC 253  & 1e5 &  30 & 1.5e17 &  3e6  & 0.08 & 0.97 & 0.30 & 0.48 & 13 & 13 \\
         & 3e3 & 150 &  3e16  & 4.5e5 & 0.92 & 0.03 & 0.70 &      &    &    \\  
NGC 613  & 1e5 &  60 &  1e17  &  6e6  & 0.15 & 0.97 & 0.23 & 0.21 & 11 & 12 \\
         & 1e3 &  30 &  6e16  &  3e4  & 0.85 & 0.03 & 0.77 &      &    &    \\  
NGC 660  & 5e4 &  60 &  1e17  &  3e6  & 0.12 & 0.84 & 0.19 & 0.15 & 13 & 16 \\
         & 1e3 &  60 &  6e16  &  6e4  & 0.88 & 0.16 & 0.81 &      &    &    \\  
NGC 1068 & 1e5 &  20 &  3e17  &  2e6  & 0.03 & 0.02 & 0.02 & 0.37 & 12 & 27 \\
         & 3e3 & 150 &  3e17  & 4.5e5 & 0.97 & 0.98 & 0.98 &      &    &    \\  
NGC 1097 & 1e5 &  60 &  6e16  &  6e6  & 0.09 & 0.87 & 0.06 & 0.16 & 10 & 13 \\   
         & 5e2 &  60 &  1e17  &  3e4  & 0.91 & 0.13 & 0.94 &      &    &    \\  
NGC 1365 & 5e4 &  60 &  1e17  &  3e6  & 0.12 & 0.84 & 0.19 & 0.15 & 13 & 16 \\ 
         & 1e3 &  60 &  6e16  &  6e4  & 0.88 & 0.16 & 0.81 &      &    &    \\  
IC 342   & 1e5 &  20 &  1e17  &  2e6  & 0.88 & 0.91 & 0.96 & 0.21 & 10 & 17 \\
         & 3e3 & 150 &  3e16  & 4.5e5 & 0.12 & 0.10 & 0.04 &      &    &    \\  
NGC 1808 & 1e5 &  60 &  3e16  &  6e6  & 0.09 & 0.67 & 0.03 & 0.28 & 13 & 33 \\
         & 1e3 & 150 &  1e17  & 1.5e5 & 0.91 & 0.33 & 0.97 &      &    &    \\  
NGC 2559 & 1e5 &  60 &  6e16  &  6e6  & 0.09 & 0.87 & 0.06 & 0.16 & 10 & 13 \\   
         & 5e2 &  60 &  1e17  &  3e4  & 0.91 & 0.13 & 0.94 &      &    &    \\  
NGC 3034 & 1e5 & 100 &  1e17  &  1e7  & 0.07 & 0.74 & 0.07 & 0.27 & 12 & 16 \\ 
         & 1e3 & 100 &  1e17  &  1e5  & 0.93 & 0.26 & 0.93 &      &    &    \\  
NGC 4945 & 1e5 &  25 &  1e17  & 2.5e6 & 0.52 & 0.92 & 0.67 & 0.20 & 16 & 20 \\
         & 1e3 & 100 &  3e16  &  1e5  & 0.48 & 0.08 & 0.33 &      &    &    \\  
NGC 5236 & 1e5 &  20 &  1e17  &  2e6  & 0.46 & 0.84 & 0.58 & 0.12 & 13 & 27 \\
         & 1e3 & 100 &  3e16  &  1e5  & 0.54 & 0.16 & 0.42 &      &    &    \\  
\noalign{\smallskip}
\hline
\end{tabular}
}%
\end{center} 
\end{table*}

The objects in this paper were all included in the previously
published survey of the $J$=1-0, $J$=2-1, $J$=3-2 transitions of $\co$
and $\thirco$ emission from galaxy centers whereas for half the sample
the $J$=4-3 $\co$ transition was also measured (Paper I). The results
of that survey were evaluated with large-velocity-gradient (LVG)
models employing the $RADEX$ radiative transfer code (Van der Tak
$\etal$ (2007). For the details of the analysis we refer to section 5
of Paper I. The LVG approximation efficiently solves the radiative
transfer equation in non-LTE environments and yields a first order
determination of the gas properties in a homogeneous medium. For each
case, the model provides an average description of all the molecular
gas in the aperture, thus lumping together gas of all temperatures and
densities. As the number of gas phases included is increased, the
models become ever more realistic. The $RADEX$ analysis,
howver, requires four input parameters per phase and per species
($\h2$ kinetic temperature and density, molecular column density per
velocity interval, relative molecular abundance). Even with
simplifying assumptions (such as identical $\h2$ temperature and
density for all species), the number of phases that can be ltaneously
modeled is severely limited by the number of independent
measurements. In the case of the low-$J$ $\co$ and $\thirco$
measurements presented in our previous work only two gas-phases can be
modeled. This allows a first, coarse separation of the dominant gas
components, such as dense or diffuse, and cool or warm. Although a
simplification of reality, this is nevertheless already a great
improvement on single-phase models that produce averages with little
physical meaning.  A complication is that for each gas phase, the
observed line intensities are always subject to degeneracies between
temperature, density, and column density per velocity interval. These
degeneracies are not always clearly resolved by the limited number of
transitions providing independent line intensities especially when
finite errors are taken into account.  Instead of a well-constrained
unique result, the two-phase modeling produced for each galaxy a
number of possible solutions. These form a well-defined and limited
range of physical parameters; examples are given in appendix C.1 of
Paper I. With only low-$J$ transitions, the number of independent
measurements is usually sufficient only to marginally constrain the
seven parameters needed to describe two phases, producing tight
constraints for some parameters but leaving others practically
unconstrained. Fits with a cut-off at the $J$=3-2 or $J$=4-3
transitions tend to underestimate the parameters of the high-pressure
gas, in particular the density. Biased fit results for the
high-pressure gas in turn influence the fit parameters of the
low-pressure gas, especially the temperature.

This is borne out by the new measurements in the $J$=6-5
transition. For each galaxy the various physical parameter sets that
provide good fits to the $J_{upp}\leq3$ intensity ratios, including
the ``best'' fits listed in Table C.2 of Paper I, fail to adequately
predict the line intensities observed in the $J$=6-5 transition. The
observed $\co$ intensities generally exceed the model-predicted values
by factors of two or more.

For $\thirco$(6-5), the result is no better.  Even in galaxies where
the predicted $22"$ model isotopologue intensity ratios are broadly
similar to the observed $9"$ APEX ratios this is only because the
individual $\co$ and $\thirco$ intensities are both off by the same
factor. Thus, the observations of the lower $J$=1-0, $J$=2-1, and
$J$=3-2 transitions alone do not sufficiently constrain the modeling
of gas physical parameters to also allow successful prediction of the
higher $J$=6-5 transition intensities. The new $J$=6-5 $\co$ and
especially $\thirco$ observations provide information on the physical
condition of the molecular gas not apparent in the lower-$J$
measurements.

This has major consequences for the modeling presented in Paper I. The
addition of two more intensities increases the number of independent
parameters to the number required to fully describe the two gas
phases. The number of possible parameter sets derived from the
$J_{upp}\leq3$ analysis for each galaxy is drastically reduced. To
agree with the $J_{upp}=6$ data, at least one of the two model phases
needs to have a kinetic temperature of 60 K or higher, and at least
one of the two phases needs to have a density of $10^4\,\cc$ or
higher. Sets falling short of this criterion can be removed -- this
includes all but a few of the sets that were earlier found to provide
possible solutions in in the analysis described in Paper I.  The
$J_{upp}\leq3$ measurements very poorly distinguish temperatures and
densities much above these values (cf. the effective upper limits in
Table\,\ref{crit}). High model temperatures and (column) densities
from the initial analysis need fine-tuning to fit the $J$=6-5 values
without compromising the low-$J$ line ratios.
    
Still assuming a [$\co$]/[$\thirco$] isotopic abundance ratio of 40
(Tang $\etal$ 2019, Viti $\etal$ 2020), we revised the two-phase
models of the sample galaxies from Paper I to accomodate the new
$J$=6-5 intensities.  The newly determined parameters of the two
phases are summarized in columns 2, 3 and 4 of
Table\,\ref{65model}. For each galaxy, two entries are given that
refer to the respective high-pressure (top) and the low-pressure phase
(bottom) identified in column 5.  Columns 6 through 8 list the
fractions of the total $\co$(1-0) and $\co$(6-5) emission and the mass
associated with each gas phase. Finally, columns 9 through 11 show the
$J$=6-5 $\co$ and $\thirco$ model line ratios for the combined
emission from the two phases. These can be compared to the observed
(6-5)/(1-0) ratios listed in Table\,\ref{corat} and the isotopologue
intensity ratios in Tables\,\ref{isotop} and \ref{litiso}. We did not
remodel NGC~6240 as it is incomparable to the other galaxies in terms
of distance, area covered, and luminosity class (see e.g., Greve
$\etal$ 2009, Papadopoulos $\etal$ 2014).

The inclusion of the $J$=6-5 measurements thus results in fits that
are much more tightly constrained than those based on the
$J_{upp}\leq4$ transitions only. This is largely due to the $\thirco$
intensities that render the isotopological intensity ratios
particularly sensitive to changes in the physical parameters. Most of
the solutions allowed by the analysis in Paper I are completely ruled
out by the present analysis. There remains a small residual
uncertainty due to the limited ability to distinguish between
temperatures above 200 K and densities well in excess of $10^{5}\cc$.

For two-phase models, much of the ambiguity previously present is
eliminated. Other parameter combinations are still possible but only
as long as they are close to those listed in Table\,\ref{65model}. It
is, however, unfeasible to assign uncertainties to individual
parameters because of the trade-offs inherent in
degeneracies. Instead, we assign very roughly a factor of two
uncertainty to the overall result. An additional source of uncertainty
is the actual $\co$/$\thirco$ abundance. Values as low as 30 and as
high as 70 have been published but most determinations settle around
40 which is the value we assumed. If in any galaxy the abundance is
different, this would lead to modestly different model parameters. We
note that such a situation seems to apply to luminous infrared
galaxies such as NGC~6240 with abundances of 100-200.

The two-phase model fits presented here provides a simplified but
robust picture of the molecular gas in the sample galaxies, especially
as concerns the division in gas of high and of low pressure. They are,
however, still a first approximation and not yet a fully realistic
description of that gas. Nevertheless, the agreement with similar
results derived independently by others is encouraging. From the
analysis of CS emission ladders in half a dozen galaxies Bayet $\etal$
(2009), for instance, conclude to the general presence of two
high-pressure phases with kinetic temperatures all below 70 K, with
similar densities 0.4-1.6 $\times$ 10$^{5}$ $\cc$ for the dominant
cold high-pressure phase but with higher densities 2.5-40 $times$
10$^{5}$ $\cc$ for the more sparse warm high-pressure phase. The
galaxy NGC~253 also provides an interesting case for comparison,
because it has been comprehensively analyzed by Rosenberg $\etal$
(2014) and P\'erez-Beaupuits $\etal$ (2018), using all available
$\co$, $\thirco$, HCN, $\ci$, and $\cii$ lines to fit three distinct
gas components. The results listed for the phases of NGC~253 in
Table\,\ref{65model} are within a factor of two of the results for the
corresponding phases in these two analyses.

Compared to our earlier analysis by Paper I, the new results in
Table\,\ref{65model} show either similar or moderately higher
temperatures and densities for the low-pressure gas forced by the new
high-pressure values. The high-pressure gas temperatures are also
roughly similar but the densities are revised up, in most cases
significantly, in order to reproduce the observed intensities of the
$\co$(6-5) and especially the $\thirco$(6-5) lines. Uncertainties are
much reduced.

The low-pressure gas is not very dense (500-3000 $\cc$) but tends to
be hot with kinetic temperatures from 60 K to 150 K. The high-pressure
gas is always very dense ($0.5-1.0\,\times\,10^{5}\cc$ or higher) and
significantly cooler with temperatures ranging from 20 K to 60 K.
Only in NGC~3034 (M~82) both gas phases have similar temperatures of
about 100 K.

Two of the twelve galaxies in Table\,\ref{65model} stand out with gas
of a single phase responsible for essentially all of their CO line
emission. Hot, moderately dense low-pressure gas produces over $95\%$
of the emission from the center of NGC~1068 independent of transition
observed, but a small amount of cold, dense gas is still required to
explain the data.  Having observed the lower transitions of HCN,
HCO$^+$ and CO isotopologues with the arcsec-sized $SMA$ and $NOEMA$
beams, Krips $\etal$ (2011) obtained a very similar result. Almost all
of the emission arises from the gas within $\sim$ 150 pc from the
active Seyfert nucleus only resolved by interferometer arrays. The
circumnuclear gas in {the starburst-dominated center of} IC~342 is
also of limited extent but here it is the high-pressure gas that
provides almost all of the CO line emission from this nearby nucleus.
The cool and very dense gas in this reservoir produces typically
$90\%$ of the CO emission again independent of observed
transition. Only ten per cent of the gas in the IC~342 nuclear region
is moderately dense but rather hot which was also concluded by
Rigopoulou $\etal$ (2013), see also Montero-Casta\~{n}o $\etal$
(2006).

In two other galaxies, NGC~4945 and NGC~5236 (M~83), both warm and
modestly dense low-pressure gas and much more dense and colder
high-pressure gas contribute in roughly equal amounts to the $J$=1-0
CO groundstate emission. Similar to IC~342, these are relatively nearby
galaxies and the line measurements sample only the gas reservoirs in
the inner few hundred parsecs.  In the likewise nearby galaxies
NGC~253 and NGC~3034 (M~82) the high-pressure gas is, however, only a
minor contributor to the groundstate CO emission as is also the case
in the other six galaxies.

Thus, the $J$=1-0 $\co$ emission from all but three of the observed
galaxy centers primarily originates in low-pressure gas reservoirs.
More than 85 per cent of the groundstate emission from these galaxies
which dominates their molecular gas mass represents moderately
dense gas at kinetic temperatures above 60 K and reaching as high as
150 K. Both temperature and density suggest a heating mechanism other
than UV and are more compatible with mechanical heating from decaying
shocks and turbulent dissipation.  Infrared spectroscopy with
$Herschel$ and $Spitzer$ already suggested this for NGC~1097
(Beir\~{a}o $\etal$ 2012).

In all galaxies except NGC~1068 the situation is completely reversed
in the $J$=6-5 transition. More than 85 per cent of the $\co$(6-5)
emission in these galaxy centers comes from relatively cool but rather
dense molecular gas reservoirs in most cases representing a minor
fraction of the total mass.

The high-pressure gas has an optical depth of a few in the $J$=6-5
transition. Although it is radiatively important, its source of
excitation is not clear-cut as various mechanisms may compete as
discussed, for instance, by Rosenberg $\etal$ (2014) for the case of
NGC~253. If the high-$J$ $\co$ emission originates in thin outer
layers of molecular clouds, it could trace high-density gas excited by
external UV radiation. The low temperature, the angular extent of the
$\co$(6-5) emission and the apparent lack of accompanying $\thirco$
emission in NGC~253 and NGC~4945 are, however, more consistent with an
extended diffuse gas excited throughout its entire volume by
mechanical heating (see Loenen $\etal$ 2008, Kazandjian $\etal$ 2015,
Paper II).

\section{Conclusion}
In this paper we have summarized all fourteen presently available sets
of $J_{upp}\geq5$ $\thirco$ measurements of galaxies beyond the Local
Group.  Together with the more abundant $\co$(6-5) measurements, they
yield thirteen $J$=6-5 $\co$/$\thirco$ isotopologue ratios for
comparison with $J_{upp}\leq3$ ratios established earlier. The
distances of the sample galaxies range from 3.5 to 21.5 Mpc. We also
observed the LIRG NGC~6240 at a distance of 116 Mpc but did not
include it in the analysis because of its discrepant nature. We have
determined $\co$(6-5) intensities in multiple apertures ranging from
$9"$ to $43"$ in ten galaxies. On average, the surface brightness of
the galaxies in this sample is roughly inversely proportional to
aperture size, indicating centrally peaked emission. The $\co$(6-5)
emission reduced to $22"$ apertures is relatively bright with
velocity-integrated (6-5)/(1-0) brightness temperature ratios ranging
from 0.12 to 0.45. A wider sample of galaxies observed in a $43"$
aperture yields on average significantly lower ratios of (6-5)/(1-0),
suggesting that the larger apertures include a higher fraction of
low-excitation gas and that molecular gas excitation increases towards
galaxy nuclei. Line intensities of $\co$(6-5) and $\thirco$(6-5) are
weakly correlated with those of HCO$^+$ and not at all with those of
HCN, although all three lines have similar (critical) densities and
presumably kinetic temperatures.

This paper not only covers all extragalactic $\thirco$(6-5)
measurements available to date but, by implication, also all available
intensity ratios $J$=6-5 $\co$/$\thirco$ that can be compared to those
of the lower $J_{upp}\leq3$ transitions. These ratios are the
emission-weighted average of a variety of molecular cloud types
ranging from dense and compact to tenuous and diffuse over relatively
large areas.  In about a third of the galaxies observed, the
isotopologue intensity ratios vary little with transition, in the
remaining two thirds the $J$=6-5 ratio is notably higher than the
ratio in the lower transitions seen in somewhat larger apertures. In
four galaxies, ratios determined in the fifteen times larger
$Herschel$ aperture increase with observed transition to much higher
values of typically 30. The increase to high values occurs around the
$J$=5-4 transition.

The actual $\co$ and $\thirco$ intensities in the $J_{upp}\geq6$
transitions are not easily predicted from the $J_{upp}\leq3$
transitions routinely available from ground based facilities. The
low-$J$ and high-$J$ lines originate in different and mostly unrelated
gas phases. Widely accessible $\co$ line intensities in the $J$=1-0
through $J$=3-2 transitions fail to fully constrain these gas phases
even when they are accompanied by complementary $\thirco$
observations. We find from our two-phase RADEX models that additional
$\thirco$ line intensities in the $J$=6-5 transition or higher
eliminate much of the degeneracy and consequent uncertainty in the
underlying physical parameters of the molecular gas. For the majority
of galaxies the models indicate that most of the observed $J$=6-5
$\co$ and $\thirco$ emission arises in a warm ($T_{kin}\geq20$K) and
very dense ($n_{\h2}\gtrsim10^{5}\,\cc$ gas.  The observed $J$=6-5 CO
emission is important as a tracer of inner galaxy energetics but not
as a tracer of inner galaxy molecular gas mass.

\begin{acknowledgements}

We gratefully acknowledge the ESO APEX User support supplied by Carlos
de Breuck. We thank Enrica Bellocchi for supplying us with the
$Herschel$-SPIRE intensities of NGC~4945 in Table\,\ref{litiso}, and
Dimitra Rigopoulou for communicating the $Herschel$-SPIRE data for
IC~342 in advance of publication.
\end{acknowledgements}


\end{document}